\documentclass[%
 reprint,
 amsmath,amssymb,
 aps,
]{revtex4-2}

\usepackage{graphicx}
\usepackage{dcolumn}
\usepackage{bm}
\usepackage{upgreek}
\usepackage{epstopdf}
\usepackage{amsmath}
\usepackage{amssymb}
\usepackage{url}

\begin{document}

\preprint{APS/123-QED}

\title{Revealing the three-component structure of water with principal component analysis (PCA) on X-ray spectrum}

\author{Zhipeng Jin$^1$, Jiangtao Zhao$^2$, Gang Chen$^3$, Guo Chen$^{4*}$, Zhenlin Luo$^{2*}$ and Lei Xu$^{1*}$}

\affiliation{$^1$ Department of Physics, The Chinese University of Hong Kong, Hong Kong, China
\\$^2$ National Synchrotron Radiation Laboratory, University of Science and Technology of China, Hefei 230026, China
\\$^3$ School of Physical Science and Technology, ShanghaiTech University, Shanghai 201210, China
\\$^4$ Chongqing Key Laboratory of Soft Condensed Matter Physics and Smart Materials, College of Physics, Chongqing University, Chongqing 400044, China}

\date{\today}
\begin{abstract}

Combining the principal component analysis (PCA) of X-ray spectrum with MD simulations, we experimentally reveal the existence of three basic components in water. These components exhibit distinct structures, densities, and temperature dependencies. Among the three, two major components correspond to the low-density liquid (LDL) and the high-density liquid (HDL) predicted by the two-component model, and the third component exhibits a unique 5-hydrogen-bond configuration with an ultra-high local density. As the temperature increases, the LDL component decreases and the HDL component increases, while the third component varies non-monotonically with a peak around 20 $^{\circ}$C to 30 $^{\circ}$C. The 3D structure of the third component is further illustrated as the uniform distribution of five hydrogen-bonded neighbors on a spherical surface. Our study reveals experimental evidence for water's unique three-component structure, which provides a fundamental basis for understanding water's special properties and anomalies.

\end{abstract}

\maketitle

\section{Introduction}
As the most common and important liquid for us, water is not common at all. Instead, it exhibits rather special properties and anomalies. For example, various properties exhibit non-monotonic behaviors, such as the maximum density at 4 $^{\circ}$C~\cite{vedamuthu1994properties,cho1997understanding,tanaka1998simple,cho1996explanation}, the minimum heat capacity at 35 $^{\circ}$C~\cite{angell1982heat}, and the minimum isothermal compressibility at 46 $^{\circ}$C~\cite{kell1970isothermal,speedy1976isothermal}, which are very different from most other liquids. These special properties originate from water's unique structure, which is still a big mystery under active debate. The main difficulty lies in the experimental front and the lack of direct experimental proof prevents a clear understanding on water's structure.

At the theoretical front, several different models and scenarios have been proposed, including the stability-limit conjecture~\cite{speedy1982stability}, the second critical point hypothesis~\cite{poole1992phase}, the critical-point-free scenario~\cite{poole1994effect,angell2008insights}, and the singularity-free scenario~\cite{sastry1996singularity}. In particular, researchers found that multiple quantities of water seem to diverge when extrapolated to $T_c=-45~^{\circ}$C~\cite{speedy1976isothermal}, which is a signature of critical point~\cite{schutz2001phase}. As a result, the second critical point hypothesis gains prevailing support ~\cite{PNAS2021,fuentevilla2006scaled,holten2012entropy,mishima1998relationship,mishima2002propagation,palmer2014metastable,debenedetti2020second}. This hypothesis proposes a liquid-liquid critical point (LLCP) for water at $T_c=-45~^{\circ}$C, below which water phase separates into the low-density liquid (LDL) and the high-density liquid (HDL), and above which water is a mixture of the two (note that the critical point also requires a pressure much higher than the atmospheric pressure and cannot be realized in normal conditions~\cite{sellberg2014ultrafast,kim2017maxima,debenedetti2020second}). Recent experiments show that water exhibits a maximum in both isothermal heat capacity~\cite{PNAS2021} and compressibility~\cite{kim2017maxima} at 1 bar, consistent with the existence of LLCP at lower temperatures (P$>$0).

However, it is very difficult to realize the critical point experimentally, because the predicted critical temperature, $T_c=-45~^{\circ}$C, is far below the freezing temperature of water and even below the homogeneous nucleation temperature at -41 $^{\circ}$C~\cite{cwilong1947sublimation,mason1958supercooling,thomas1952889}. Consequently, water will spontaneously crystallize far above $T_c$ and it is very difficult to observe LDL and HDL in their bulk phases and their transition until a recent experiment performed at very rapid speed and high pressures \cite{kim2020experimental}. Although crystallization can be inhibited through strong confinement~\cite{cerveny2016confined,liu2005pressure,gallo2016water}, adding solutes ~\cite{corradini2010effect,woutersen2018liquid} or some other special treatments~\cite{kim2020experimental,kringle2020reversible,sellberg2014ultrafast}, the results under such special conditions may not be representative for pure bulk water at normal pressure. As a result, it is more practical to numerically probe the components of water with molecular dynamics(MD) simulations~\cite{debenedetti2020second,moore2011structural,palmer2014metastable,tanaka1996self,smallenburg2014erasing,xu2006thermodynamics}, which however give very different conclusions with different water models. Therefore, an objective and systematic experimental analysis, which can extract direct and unambiguous experimental evidence, is currently the most critical issue in this research field.

\begin{figure*}[htpb]
\includegraphics[width=17.2cm]{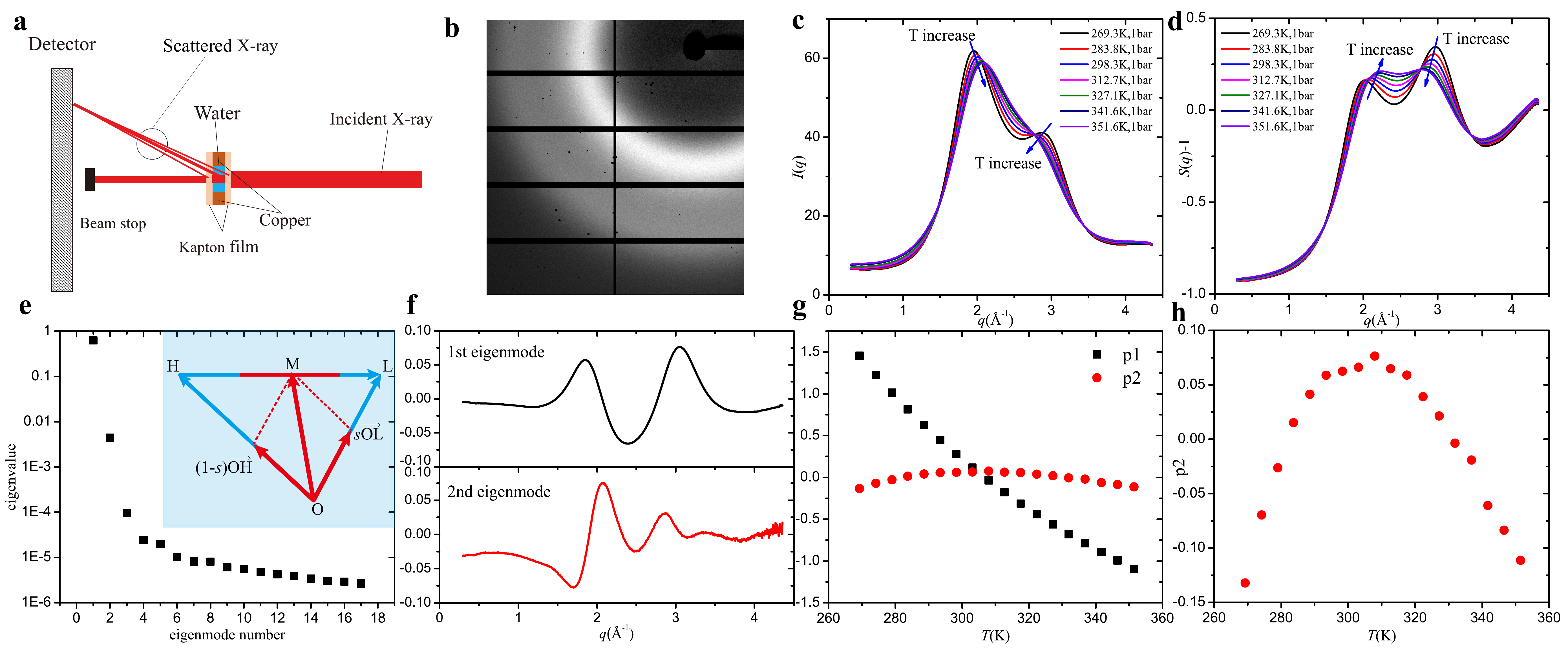}
\caption{\small\label{fig:1}(a) The schematics of experimental setup. (b) A typical X-ray scattering image. The horizontal and vertical dark regions are gap regions without active pixels. (c) Scattering intensity $I(q)$ under different temperatures (-5 $^{\circ}$C to 80 $^{\circ}$C). (d) Structure factor $S(q)$ under different temperatures. (e) Main panel: eigenvalues from large to small. The first two are significantly larger than the rest. Inset: vector analysis of the first eigenmode. $\protect\overrightarrow{OL}$, $\protect\overrightarrow{OH}$ and $\protect\overrightarrow{HL}$ represent LDL, HDL, and the first eigenmode respectively. $\protect\overrightarrow{OM}$ represents an actual sample with a fraction of $s$ low-density liquid (LDL) and $(1-s)$ high-density liquid (HDL). According to vector addition, point $M$ locates on the straight line of $\protect\overrightarrow{HL}$: $\protect\overrightarrow{OM}=s\protect\overrightarrow{OL}+(1-s)\protect\overrightarrow{OH}=s(\protect\overrightarrow{OL}-\protect\overrightarrow{OH})+\protect\overrightarrow{OH}$, which gives $\protect\overrightarrow{OM}-\protect\overrightarrow{OH}=s(\protect\overrightarrow{OL}-\protect\overrightarrow{OH})$ and thus $\protect\overrightarrow{HM}=s\protect\overrightarrow{HL}$. As $s$ varies with temperature, point $M$ sweeps through the red segment on $\protect\overrightarrow{HL}$ and each point on this red segment corresponds to one actual sample we can experimentally measure. (f) The first and second eigenmodes calculated from $S(q)$ curves. (g) Each $S(q)$ curve measured in (d) is projected onto the first and second eigenmodes with $p1$ and $p2$ the projection pre-factors. $p1$ decreases significantly while $p2$ is quite stable. (h) Zoomed-in plot of $p2$ shows a non-monotonic behavior.}
\end{figure*}

\section{X-ray diffraction experiment}
To address this critical issue, we turn to a powerful mathematical tool, the principal component analysis (PCA). It is a mature and robust approach for identifying the crucial components or dominant factors from their complex combinations. We apply this powerful analysis onto X-ray spectrum of liquid water at different temperatures (-5 $^\circ$C to 80 $^\circ$C), and combine it with numerical simulations to illustrate the microscopic structure of water. As expected, the experimental data unambiguously reveal two major components with LDL-like and HDL-like structures, which convert into each other as temperature changes. More strikingly, our data further uncover a third component, which exhibits distinct structure, density, and temperature dependency from the other two. Combining experiment with simulation, our study elucidates the unique three-component structure of water, and provides a fundamental basis to understand water's special properties and anomalies.

To probe water's microscopic structure, we perform X-ray scattering experiment in the BL19U2 station of Shanghai Synchrotron Radiation Facility (SSRF). Our X-ray beam has the spot size of 320*43 $\upmu$m$^{2}$ and the energy of 12.000 $\pm$ 0.002 keV. After incidenting onto the deionized water sample, the scattered X-ray is collected by a Pilatus1M area detector. The schematics of the setup is shown in Fig.~1(a) and a typical raw scattering image is shown in Fig.~1(b). To maximize the measurement range in $q$, the detector only collects about one-quarter of the scattered light, as shown in Fig.~1(b). The temperature of the deionized water is systematically varied by a thermal stage (Linkam HFSX350) between -5 $^{\circ}$C to 80 $^{\circ}$C, and we record the X-ray scattering image after the sample temperature is stabilized at designated values. The water temperature is measured in real time, by a Micro-BetaCHIP Thermistor probe placed adjacent to the incident area (within 0.5 mm). More experimental details can be found in the Appendix.

After the data processing of image integration, background subtraction, absorption correction, geometry and polarization correction (see details in Appendix), we obtain the scattering intensity curves $I(q)$ under different temperatures, as plotted in Fig.~1(c). The structure factor $S(q)$ can be further obtained from $I(q)$ (see details in Appendix), as plotted in Fig.~1(d), which is consistent with previous measurements \cite{JCP2014}. Based on these $S(q)$ curves, we perform the principal component analysis (PCA).

\section{Principal component analysis}
PCA is a powerful mathematical approach (see Appendix A) widely used in various fields, such as statistics, data science, quantitative finance, neuroscience, and physics~\cite{tan2012understanding,shen2016probing}. Analogous to finding the principal axes of a rigid body, PCA identifies the most critical eigenmodes of the covariance matrix built from different measurements, and illustrates the data with these critical modes. For a complex system composed by multiple components, the eigenmodes of PCA can reveal the valuable information of basic components, and the projections along these eigenmodes can provide the component weight information. Compared with the previous analysis on water structure~\cite{russo2014understanding,shi2020direct}, our PCA approach has the advantage of being completely objective without any fitting, systematically combining all data from different measurements, and thus can provide unambiguous and robust experimental evidence.

Based on the different $S(q)$ measurements in Fig.~1(d), we construct the covariance matrix and obtain its eigenvalues and eigenmodes (see Appendix). Fig.~1(e) main panel shows the eigenvalues arranged from large to small: apparently, the first and second eigenvalues are significantly larger than the rest, which are small and at noise level. Thus most physics in our data can be represented by the first two eigenmodes, which indicate two main reasons for water structure evolution under different temperatures. We will show that the first reason is due to the mutual conversion between two dominant water components, and the second reason is related to the third component of water. Note that PCA on $I(q)$ curves also give the same results, as shown in Appendix.

\begin{figure*}[htpb]
\includegraphics[width=17.2cm]{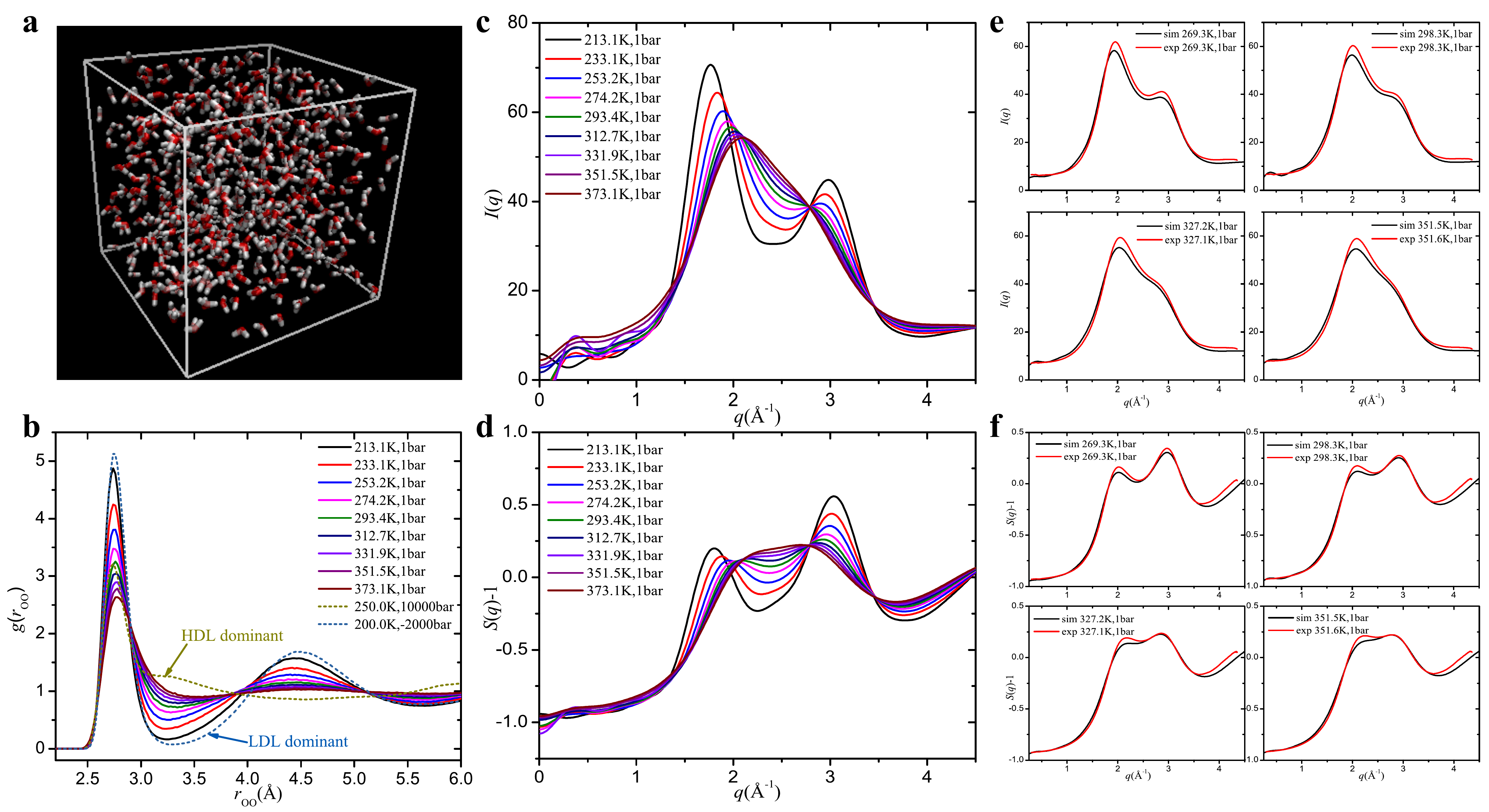}
\caption{\small\label{fig:2}(a) A snapshot of 512 water molecules at 25 $^{\circ}$C simulated by TIP4P-Ew model. We average over 2,000 snapshots to obtain simulation data. (b) The radial distribution function of oxygen atoms simulated at different temperatures. The LDL and HDL dominant systems are shown by dashed curves. (c) The simulated scattering intensity $I(q)$ at different conditions. (d) The simulated structure factor $S(q)$ at different conditions. (e) and (f) The comparison of $I(q)$ and $S(q)$ between simulation and experiment at 4 different temperatures.}
\end{figure*}

The first eigenvalue accounts for the majority of the total sum, indicating that the physics is dominated by the first eigenmode and thus essentially one-dimensional (1D) (note that the eigenvalues are the variance of raw data projected along each eigenmode but not the fraction of each component). This 1D behavior is consistent with a mutual conversion between two basic components as proved below. Without loss of generality, the basic components and their superpositions can be represented by vectors in the eigenmode space. Therefore, we use two vectors, $\overrightarrow{OH}$ for HDL and $\overrightarrow{OL}$ for LDL, to represent the two basic components of water in Fig.~1(e) inset. Linear combinations of $\overrightarrow{OH}$ and $\overrightarrow{OL}$, which represent mixtures of HDL and LDL, can in general spread in a 2D plane. However, the mutual conversion means that one changes into another but their sum remains as a constant. This constraint eliminates one degree of freedom and collapses the 2D combinations into a 1D straight line: assuming that an experimental system contains LDL with a fraction of $s$ and HDL with a fraction of $(1-s)$, its vector representation, $\overrightarrow{OM}=s\overrightarrow{OL}+(1-s)\overrightarrow{OH}$, must lie on the straight line of vector $\overrightarrow{HL}$ for an arbitrary $s$, as shown by the vector analysis in Fig.~1(e) inset. As we vary the temperature from -5 $^\circ$C to 80 $^\circ$C, the fraction $s$ will change correspondingly and $\overrightarrow{OM}$ will sweep across the red segment on $\overrightarrow{HL}$. Therefore, if water is indeed mainly composed by two components, their mutual conversion should generate 1D behavior with one dominant eigenvalue and eigenmode, consistent with our observation. Furthermore, this dominant eigenmode has the physical meaning of vector $\overrightarrow{HL}$ (after renormalization), which means the difference between LDL and HDL: $\overrightarrow{HL}=\overrightarrow{OL}-\overrightarrow{OH}$. We plot this first eigenmode in Fig.~1(f) upper panel and its physical meaning as vector $\overrightarrow{HL}$ will be experimentally confirmed later.

Besides the dominant first eigenvalue, the second eigenvalue also exceeds the rest by a significant amount, indicating new physics beyond the two-component picture. The second eigenmode is plotted in Fig.~1(f) lower panel, which exhibits more characteristic peaks and valleys than the first mode. Because the second eigenmode is orthogonal to the first, it must be independent of the two components' conversion and indicates the existence of a third component we will verify later. We emphasize that despite the discovery of the third component, the two-component model does describe the major water structure, and the third component is a small yet significant ingredient. Its discovery demonstrates the high sensitivity of our PCA analysis for picking up small yet significant signals.

To understand the water structure based on the first and second eigenmodes, we project the measured $S(q)$ curves onto the two modes with the projection prefactors $p1$ and $p2$, as shown in Fig.~1(g). Later we will show that $p1$ and $p2$ correlate to the various components' weights in the actual samples. Clearly, $p1$ reduces significantly with $T$, corresponding to the decrease of LDL component as verified later. For the second eigenmode, $p2$ is relatively stable and non-monotonic, suggesting a distinct behavior of the third component. An enlarged picture for $p2$ is plotted in Fig.~1(h), which reveals a peak around 30 $^\circ$C. To summarize, PCA reveals intriguing clues from experimental data; however a complete understanding requires molecular level information, which will be achieved by combining experimental data with molecular dynamics (MD) simulations.

\begin{figure*}[htpb]
\includegraphics[width=17.2cm]{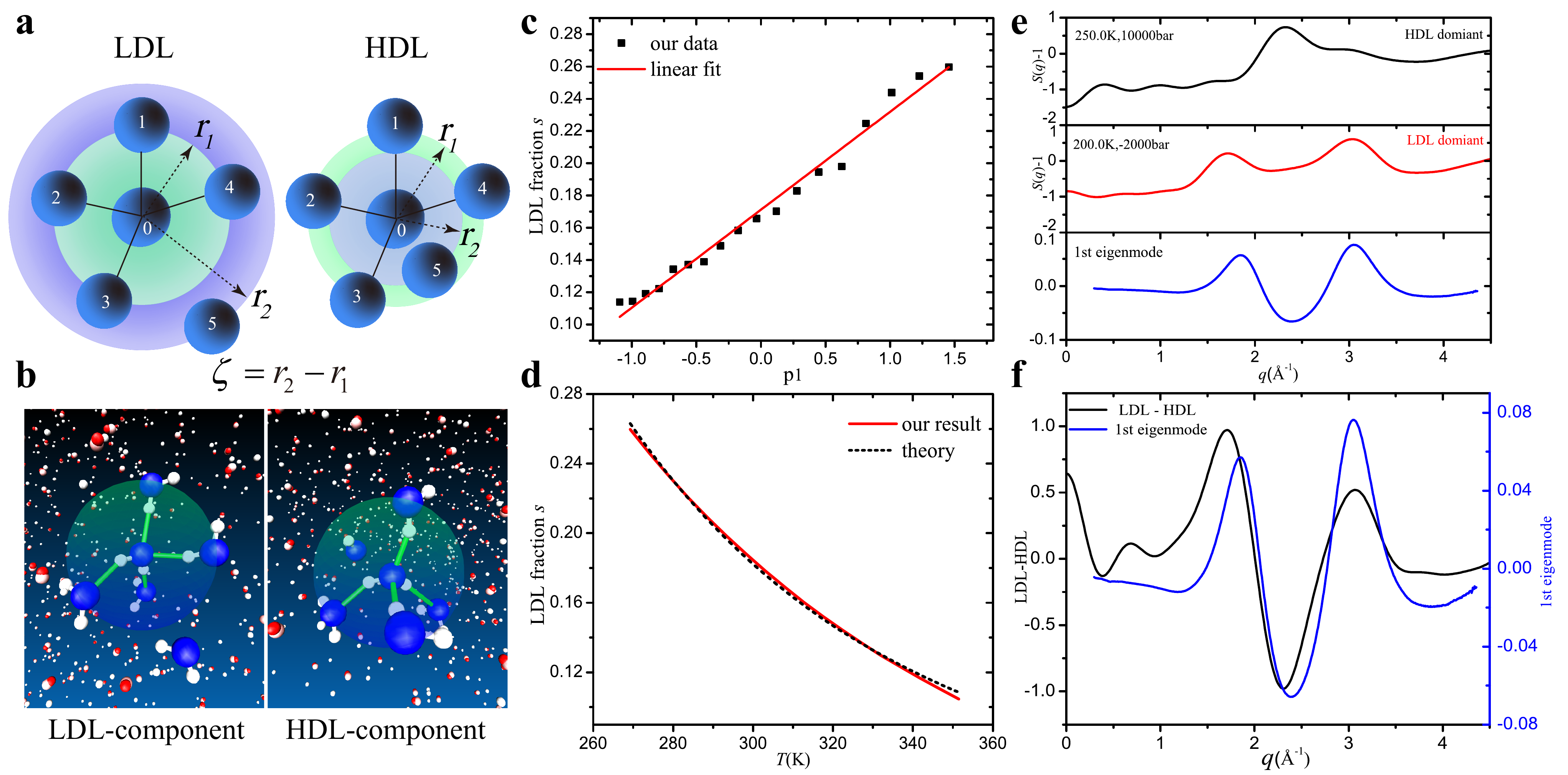}
\caption{\small\label{fig:3}(a) Schematics illustrating LDL and HDL. Molecules 1 to 4 are hydrogen bonded to 0 while 5 is not. $r_1$ is the distance to the furthermost hydrogen-bonded molecule, and $r_2$ is the distance to the closest non-hydrogen-bonded molecule. Following the previous research \cite{russo2014understanding}, the order parameter is defined as: $\zeta=r_2-r_1$. For LDL, $r_2$ is well beyond the $r_1$ shell, while for HDL $r_2$ is very close to or even enters the $r_1$ shell. (b) Typical LDL and HDL components observed in our simulation. The relevant molecules are enlarged and shown in blue, and the hydrogen bonds are indicated by the green connections. (c) LDL fraction has a linear relation with the projection pre-factor $p1$. (d) Our result agrees well with the theoretical fitting. (e) Top to bottom panels show HDL system's $S(q)$, LDL system's $S(q)$ and the first eigenmode. The locations of the three main features in the eigenmode agree well with the HDL and LDL features: the two peaks correspond to the two LDL peaks, and the valley corresponds to the HDL peak. (f) The first eigenmode agrees well with the curve obtained by the LDL curve minus the HDL curve.}
\end{figure*}
\section{molecular dynamics simulations and two-component model}
We simulate with multiple popular water models, including SPC/E, TIP3P, TIP4P, TIP4P-Ew and TIP5P, and find that the TIP4P-Ew model fits our experiment the best and will be used throughout this work (see more simulation details in Appendix). A snapshot of 512 water molecules equilibrating at 25 $^{\circ}$C is shown in Fig.~2(a), and we average over 2,000 such snapshots to achieve good statistics. With the locations of every water molecule obtained from simulations, we plot the radial distribution function of water molecules, $g(r_{OO})$, under various temperatures in Fig.~2(b). For comparison, we also simulate the LDL-dominant and HDL-dominant samples based on the previous literature~\cite{russo2014understanding,shi2020direct}, and plot them as dashed curves. Due to the small system size and short time scale, the simulated samples can reach down to -60 $^\circ$C without freezing, and clearly the structure approaches LDL as the temperature lowers. After Fourier transform, we can further obtain the scattering intensity $I(q)$ and the structure factor $S(q)$ (see Appendix for details), as plotted in Fig.~2(c) and (d). Except a much broader temperature range explored, the simulations exhibit a very similar trend as the actual measurements in Fig.~1(c) and (d). Fig.~2(e) and (f) demonstrate the direct comparison between simulation and experiment at 4 different temperatures, and the agreement between experiment and simulation is excellent. Therefore, the water structures obtained in the simulations can be considered as a reliable representation of the actual samples in the experiment.

Next we verify the two-component model which was well established theoretically but having very rare direct experimental proof\cite{shi2020direct}. Based on the previous research~\cite{russo2014understanding}, we distinguish the high-density liquid (HDL) and the low-density liquid (LDL) with the order parameter, $\zeta=r_2-r_1$: for an arbitrary water molecule, $r_1$ is the distance to its furthermost hydrogen-bonded neighbor and $r_2$ is the distance to its closest non-hydrogen-bonded neighbor, as shown in Fig.~3(a). This definition differentiates LDL and HDL by comparing the hydrogen-bonded neighbors with the non-hydrogen-bonded ones. For the case of LDL, the hydrogen-bonded neighbors (typically 4 of them) form the nearest shell within $r_1$ while the non-hydrogen-bonded neighbors are far away from this shell, i.e., $\zeta>0$ significantly. In the HDL situation, however, the non-hydrogen-bonded neighbors are very close to or even enter into this $r_1$ shell, producing a larger local density as shown in Fig.~3(a) right panel. These two components are indeed commonly observed in our simulations, as shown in Fig.~3(b): the relevant molecules are enlarged and colored in blue, and the hydrogen bonds are drawn as green long connections. Clearly, for LDL the closest non-hydrogen-bonded neighbor is far away from the $r_1$ shell while for HDL it is inside the $r_1$ shell.

Based on this definition of LDL and HDL, we can find their fractions at different temperatures in simulations (see Appendix). We then combine the fraction information from simulation with experimental data, and verify the two-component model experimentally. According to the model, any actual system we can experimentally measure is a mixture of LDL and HDL: $\overrightarrow{OM}=s\overrightarrow{OL}+(1-s)\overrightarrow{OH}=s(\overrightarrow{OL}-\overrightarrow{OH})+\overrightarrow{OH}$, with $s$ and $(1-s)$ the fractions of LDL and HDL, and vectors $\overrightarrow{OM}$, $\overrightarrow{OL}$ and $\overrightarrow{OH}$ representing the actual mixture system, LDL and HDL respectively (see Fig.~1(e)). According to this vector equation, $\overrightarrow{OM}$ depends linearly on $s$ and so does point $M$'s coordinate on $\overrightarrow{HL}$. Because $M$'s coordinate on $\overrightarrow{HL}$ can be represented by the projection prfactor $p1$, the two-component model thus predicts a linear relation between $p1$ and $s$. To test it, we plot $s$ versus $p1$ in Fig.~3(c) and obtain a very nice linear dependence. As $p1$ and $s$ are independently obtained from experiment and simulation, this linear relation shows a clear evidence for the two-component model, as well as a good agreement between our experiment and simulation.

We further compare this essential result, i.e., the linear line in Fig.~3(c), with the theory. Because all samples, both in simulation and experiment, are measured under specific temperatures, $T$ is thus a hidden variable in Fig.~3(c). Therefore, we can change variables and plot the straight line of Fig.~3(c) in terms of the variables $s$ and $T$, as shown in Fig.~3(d). For comparison, the theoretical prediction from two-component model ~\cite{tanaka2000simple,russo2014understanding,shi2018common} is also plotted as the dashed curve, which agrees excellently with our result. According to this theory, we have the following relation between the LDL fraction $s$ and the temperature $T$:
\begin{equation}
s = \frac{1}{{1 + {e^{\frac{{\Delta E}}{{{k_B}T}} - \frac{{\Delta \sigma }}{{{k_B}}}}}}}
\label{eq:1}
\end{equation}
\noindent where $\Delta E{\rm{ = }}{E_{LDL}} - {E_{HDL} }$ and $\Delta \sigma {\rm{ = }}{\sigma _{LDL}} - {\sigma _{HDL} }$ are the energy and entropy difference between LDL and HDL respectively. By fitting the theoretical dashed curve to our solid curve, we get: $\Delta E/k_B=-1238.3$ K and $\Delta \sigma/k_B=-5.63$, which are consistent with previous study~\cite{shi2020direct}. Apparently, LDL has a lower energy and entropy than HDL, because LDL typically exhibits four hydrogen bonds per molecule (3.98 in our simulation) with a relatively-ordered structure, while HDL contains a mixture of three and four hydrogen bonds (averaged at 3.79 in our simulation) with a more disordered structure. As a result, LDL is perferred at low temperatures due to its lower energy, while HDL is preferred at high temperatures due to its higher entropy. Moreover, the Schottky temperature~\cite{shi2018common} at which LDL and HDL components have equal amount, $T(s=0.5)=\Delta E/\Delta \sigma=220.0$ K (i.e., -53 $^\circ$C), is close to the critical two-phase separation temperature 228.1 K (i.e., -45 $^\circ$C), as we naturally expect. Also the relationship of Eq.~(1) works nicely in simulations over a much broader temperature range (-60 $^{\circ}$C to 100 $^{\circ}$C) than our experiment, as shown in the Appendix.

Besides the theoretical agreement, we further illustrate the physical meaning of the first eigenmode, as promised earlier. According to the vector analysis in Fig.~1(e), the first eigenmode should correspond to the vector $\overrightarrow{HL}$ (after renormalization), which is the difference between LDL and HDL: $\overrightarrow{HL}=\overrightarrow{OL}-\overrightarrow{OH}$. We directly test it in $q$ space, by comparing the first eigenmode from experiment with the difference of LDL and HDL from simulation. We first numerically construct HDL and LDL dominant systems under special conditions: the HDL-dominant ($97.7\%$) system is realized at very high pressure ($P=10000$ bar, $T=250.0$ K), and the LDL-dominant ($72.3\%$) system is achieved at very low pressure and temperature ($P=-2000$ bar, $T=200.0$ K). Both systems agree well with previous studies~\cite{soper2000structures,kim2020experimental}. Due to the very short time scale of simulations (10 to 20 ns), the LDL system can remain as liquid even at very low temperature. The HDL and LDL dominant systems are plotted in the top two panels of Fig.~3(e), and the first eigenmode is plotted in the bottom panel for direct comparison.

Apparently, there are three main features in the bottom panel (2 peaks and 1 valley), whose locations or characteristic sizes match nicely with the main features of the top two panels (2 peaks in LDL and 1 peak in HDL). We then directly subtract the top two curves, i.e., LDL minus HDL, and compare this difference against the first eigenmode, as shown in Fig.~3(f). The main features and trend match very well, confirming the picture that the first eigenmode corresponds to the difference between LDL and HDL. Once again, two independent data sets, the eigenmode from experiment and the LDL minus HDL curve from simulation, agree well with each other. This provides another strong experimental evidence for the two-component model.

\begin{figure*}[htpb]
\includegraphics[width=17.2cm]{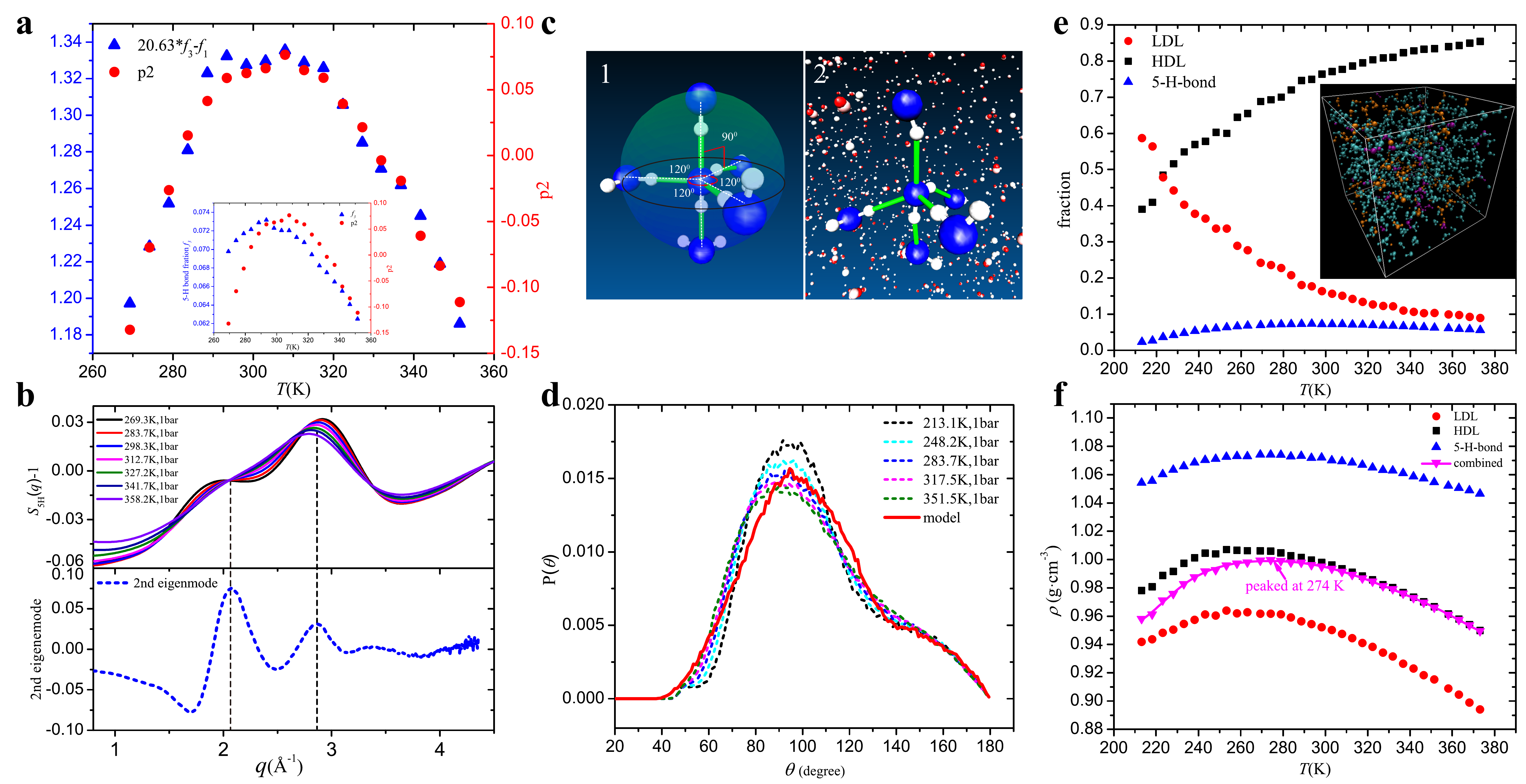}
\caption{\small\label{fig:4}(a) Inset: the 5-H-bond fraction $f_3$ from simulation and the projection pre-factor $p2$ from experiment are both non-monotonic and resemble each other. Main panel: the non-monotonic $p2$ matches $20.63\times f_3-f_1$ nicely. (b) The main peaks in the second eigenmode (bottom panel) at $q=2.06~\mathrm{\AA^{-1}}$ and $q=2.88~\mathrm{\AA^{-1}}$ correspond exactly to the isosbestic point and the main peak of $S_{5H}(q)$ (top panel). $S_{5H}(q)$ is the structure factor calculated from 5-H-bond molecules. (c) Left panel: schematics showing the uniform distribution of five hydrogen-bonded molecules on a spherical surface, with two molecules at the north and south poles, and three on the equator with $120^\circ$ angle between them. Right panel, a typical 5-H-bond structure in an actual simulation exhibits a very similar configuration as the left panel, differing only by some thermal distortions. (d) The distribution of angles between 5-H-bond oxygen atoms, $\mathrm{P(\theta)}$. The solid curve from our uniform distribution model plus random fluctuations (mimicing thermal fluctuations) agrees well with the actual simulations at different temperatures (dashed curves). (e) Main panel: the fractions of three components at various $T$. Inset: a snapshot at 293K showing a perfect mixing of three components at molecular level.  orange: LDL, cyan: HDL, purple: 5H-bond. (f) The densities of the three components and their combined density. The 5-H-bond component is significantly denser than both HDL and LDL. The combined density is peaked at $274$ K that is close to the density maximum of water at $277$ K.}
\end{figure*}
\section{The third component and the three-component picture of water}
More interestingly, our study goes beyond the two-component picture and reveals the experimental evidence for the third component. According to our PCA analysis, besides the dominant first eigenvalue, the second eigenvalue is also significantly larger than the rest and contains important physical information (see Fig.~1(e)). Because different PCA modes are orthogonal to each other, the second mode must be independent of the first, and reveals valuable information beyond the two-component picture. Previous study has uncovered some clues about the existence of a third component~\cite{segtnan2001studies}, and pioneering numerical study has proposed a possible candidate with five hydrogen bonds (5-H-bond)~\cite{yang2019ultra}. We now reveal direct experimental evidence for its existence and illustrate the 3D structure.

We again illustrate the third component with independent data sets from experiment and simulation: the projection prefactors along the second eigenmode, $p2$, are obtained experimentally, and the fractions of 5-H-bond component under different temperatures, $f_3$, are obtained numerically. The two sets of data are plotted together in Fig.~4(a) inset and their profiles resemble each other. Moreover, further analysis (shown later in Fig.~5(h) and Appendix A) indicates that $p2$ linearly correlates to the fractions of both the first component $f_1$ (i.e., LDL fraction) and the third component $f_3$ (i.e., 5-H-bond fraction): $p2 \propto 20.63\times f_3-f_1$. In the main panel of Fig.~4(a) we plot them together and observe an almost precise match. Such an excellent agreement unambiguously proves that the third component is indeed the 5-H-bond structure.

To get even more evidence, we compare the characteristic sizes in $q$ space. Due to the lack of knowledge on the formation conditions, we cannot construct a system dominated by the 5-H-bond component like LDL or HDL dominant system. However, we can extract the 5-H-bond information from normal systems: we only pick out the molecules surrounded by five hydrogen bonds, and set such molecules as origin to calculate the radial distribution function, $g_{5H}(r)$. We then Fourier transform $g_{5H}(r)$ to get $S_{5H}(q)$, and compare it with the second eigenmode from experiment, as shown in Fig.~4(b). Clearly, there are two pronounced peaks in the second eigenmode curve at $q=2.06~\mathrm{\AA^{-1}}$ and $q=2.88~\mathrm{\AA^{-1}}$, which correspond exactly to the isosbestic point~\cite{segtnan2001studies} and the main peak of the $S_{5H}(q)$ curves in the upper panel. This nice match once again confirms the 5-H-bond structure as the third component. In addition to peaks, there are also three major valleys in the second eigenmode, which come from LDL and HDL as they are close to the three main peaks in LDL and HDL (see Fig.~3(e)) (Although the second eigenmode is orthogonal to the \emph{difference} of LDL and HDL, it still contains features from LDL and HDL).

Next we illustrate the 3D structure of the third component, i.e., how do the five hydrogen bonds locate spatially. Because the five hydrogen bonds have similar lengths, we can approximate the five hydrogen-bonded molecules as locating on a spherical surface. We consider the most uniform distribution under which the molecules are separated to the largest degree. As shown in Fig.~4(c) left panel: two molecules locate on the north and south poles respectively, and three on the equator with $120^\circ$ angle between their connections. Indeed, typical 5-H-bond structures in our simulations exhibit very similar configurations as this assumption (except with some thermal fluctuations), as shown in the right panel of Fig.~4(c).

To statistically verify this most uniform distribution model, we collect a large amount of 5-H-bond molecules from actual simulations, connect their oxygen atoms, and statistically calculate the distribution of angles between these connections, $\mathrm{P(\theta)}$. For comparison, we also calculate $\mathrm{P(\theta)}$ generated from our uniform distribution model. In the ideal situation, there are only three possible angles: $90^\circ$ ($60\%$), $120^\circ$ ($30\%$), and $180^\circ$ ($10\%$) (left panel of Fig.~4(c)), which can however broaden into a continuous distribution after random perturbations that mimic thermal fluctuations (see Appendix). The curves from actual simulation (dashed) and our model calculation (solid) are compared in Fig.~4(d) and an excellent agreement is observed. This agreement confirms that the 3D structure of the third component is the most uniform distribution of five hydrogen-bonded molecules around the center, as demonstrated in Fig.~4(c).

To illustrate the overall three-component picture, we plot their individual fractions in Fig.~4(e) which add up to unity. As temperature increases, LDL decreases and HDL increases significantly, as the result of their mutual conversion. However, the third component with 5-H-bond is much more stable, with a distinct non-monotonic behavior peaked around $300$ K. This indicates a fundamentally different thermal response of the third component. In addition, the fraction of the third component is around $5\%$ to $7\%$, which is significant although the other two components are more important. Correspondingly, the third component causes a few percent modifications to the two-component model and the theoretical predictions of Eq.~(1), which are shown in the Appendix. The inset shows a snapshot at $293K$ with a perfect mixing of three components at molecular level. Our study thus provides robust evidence for the two-component model, and at the same time reveals valuable information for the third component.

We further calculate the densities of the three components with Voronoi cells obtained from each molecule, as shown in Fig.~4(f) (see Appendix for details). Interestingly, the third component exhibits an ultrahigh density that is significantly higher than both HDL and LDL, agreeing with the previous study~\cite{yang2019ultra}. When the three densities are combined according to their fractions, $\frac{1}{\rho}=\frac{f_{1}}{\rho_{1}}+\frac{f_{2}}{\rho_{2}}+\frac{f_{3}}{\rho_{3}}$, we obtain the bulk water density $\rho$ from molecular level. As shown by the connected data curve in Fig.~4(f), $\rho$ has a peak at $274$ K, which is close to the maximum water density at $277$ K. Therefore, our three-component picture at microscopic level agrees with the macroscopic density anomaly of bulk water.

\begin{figure*}[htpb]
\includegraphics[width=17.2cm]{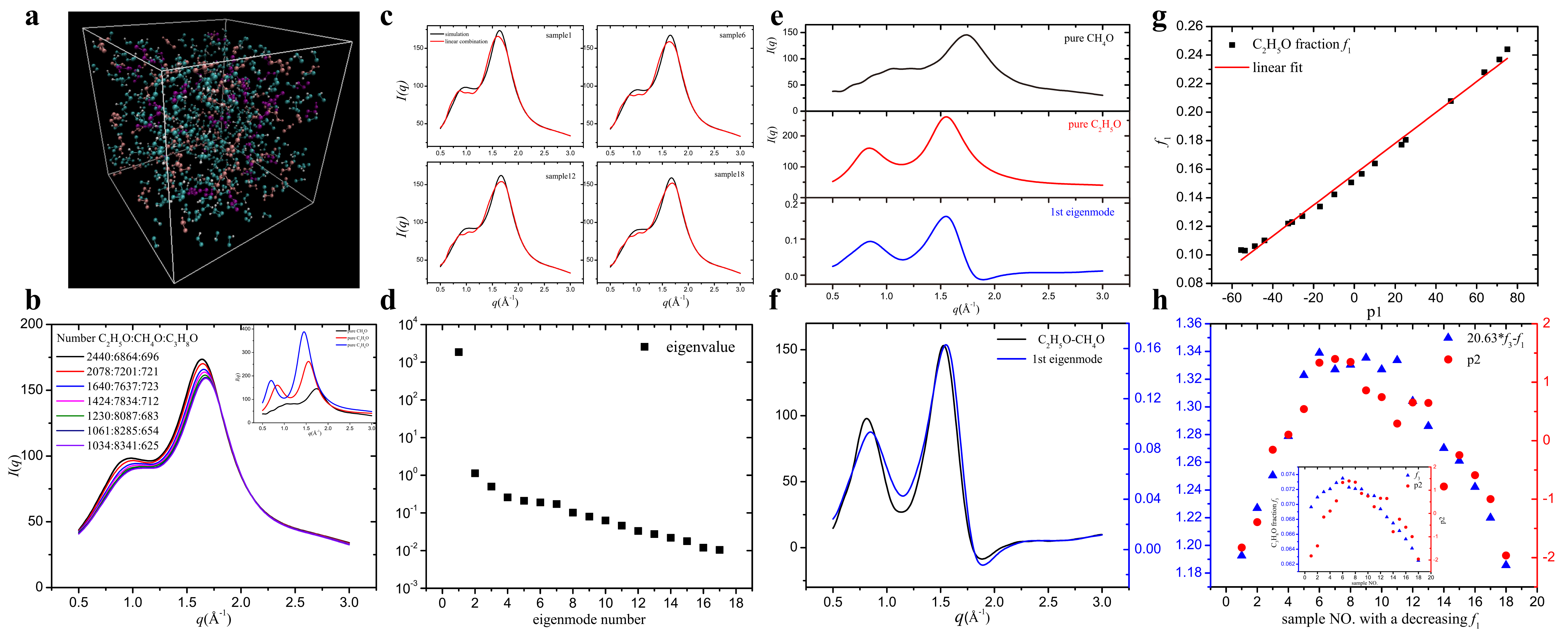}
\caption{\small\label{fig:5}(a) A snapshot of 512 alcohol molecules ($5\%$ of total system) at 25 $^{\circ}$C simulated with GROMOS 54A7 force field. The cyan, orange and purple represent CH$_4$O (methanol), C$_2$H$_6$O (ethanol) and C$_3$H$_8$O (1-propanol) respectively. (b) Several typical scattering intensity curves, $I(q)$, of the system with different fractions of the three alcohols. Inset: $I(q)$ curves for the three pure alcohols. (c) In a mixture, the linear superposition of three pure alcohols' $I(q)$ based on their fractions agree well with the mixture's actual $I(q)$. (d) The eigenvalues arranged from large to small. (e) Bottom panel: the 1st eigenmode from PCA analysis on $I(q)$. Middle panel: pure ethanol's $I(q)$. Top panel: pure methanol's $I(q)$.(f) The comparison between the difference of the two dominant components and the 1st eigenmode. They agree well with each other. (g) $p1$ has a nice linear relationship with ethanol's fraction $f_1$. (h) Inset: non-monotonic profiles of $p2$ and $f_3$ resemble each other, indicating their close correlation. Main panel: $p2 \propto 20.63\times f_3-f_1$  is verified excellently.}
\end{figure*}

\section{Comparison with the three-alcohol system}
To ensure that our PCA method reliably reveals the three components of water, and more importantly to establish it as a general approach to distinguish the components of various mixture systems, we precisely design a well-defined three-component system and check whether it reproduces all the PCA results of water. We construct a distinct system with three types of alcohols: C$_2$H$_6$O (ethanol, 0.789 g/mL), CH$_4$O (methanol, 0.792 g/mL), and C$_3$H$_8$O (1-propanol, 0.804 g/mL), which mimic water's LDL, HDL and 5-H-bond structures respectively according to their low-to-high densities. Previously our water sample is measured at 18 different temperatures and thus 18 combinations of its three components' fractions are obtained in Fig.~4(e); correspondingly, 18 mixtures of the three alcohols with identical molar fractions are numerically constructed. These precisely-designed mixtures are then analyzed with PCA and compared against the water's results. In these MD simulations, 10,000 molecules in total are put inside a cubic box, and the mixtures reach equilibrium at 25 $^{\circ}$C and 1 bar. One example is shown in Fig.~5(a), which shows a perfect mixing of three alcohols just like water's three components. Several typical scattering intensity curves, $I(q)$, are shown in Fig.~5(b) and the inset shows the $I(q)$ curves of three pure alcohols.

First we use this system to address a fundamental open issue: the vector analysis in Fig.~1(e) assumes the mixture's signal as a linear superposition of the pure components' signals, which has not been verified. With the three-alcohol system whose components' fractions are precisely known, we check this fundamental assumption directly. For any mixture we can linearly add up its three pure components' $I(q)$ based on their fractions and compare the sum with the mixture's actual $I(q)$, as shown in Fig.~5(c): the excellent match in several typical samples unambiguously proves this basic linear assumption as a good approximation. Thus all our previous linear analysis are fundamentally self-consistent.

Next we perform PCA analysis on the three-alcohol system. We directly analyze the $I(q)$ instead of $S(q)$ curves because the three alcohols have different molecular form factors and thus the mixtures only have $I(q)$ but not $S(q)$. In water we PCA analyze $S(q)$ curves because of the convenient comparison with previous studies that frequently use $S(q)$; however the PCA analysis on $I(q)$ is more general and valid for both water and other mixture systems\cite{jacs2009}. The analysis on water's $I(q)$ and $S(q)$ also give the same results (see Appendix) and thus they are essentially equivalent.

All the water's PCA results are reproduced in the three-alcohol system: there is a dominant first eigenmode correlated to the fraction variations between the two major components, CH$_4$O and C$_2$H$_6$O, and a second eigenmode correlated to the third component, C$_3$H$_8$O (see Fig.~5(d) for eigenvalues). In Fig.~5(e) bottom panel the first eigenmode exhibits two peaks and one valley which correspond to the main features in C$_2$H$_6$O (middle) and CH$_4$O (top) respectively, and in Fig.~5(f) the first mode agrees well with the two curves' difference, C$_2$H$_6$O - CH$_4$O (similar to LDL-HDL in Fig.~3e and f).  For mode projections, the projection pre-factor on the first mode, $p1$, again exhibits an excellent linear dependence with the first component's fraction $f_1$, as shown in Fig.~5(g). For the second mode, $p2$ also has a similar non-monotonic profile as the third component's fraction $f_3$, indicating their close correlation (see Fig.~5(h) inset). Theoretical analysis further finds that $p2$ directly correlates to the linear combination of $f_3$ and $f_1$: $p2 \propto 20.63\times f_3-f_1$ (see Appendix A), and the two data sets match each other very well as shown in Fig.~5(h). Moreover, this exact expression also works excellently in the water system, as shown in Fig.~4(a). One identical expression, $p2 \propto 20.63\times f_3-f_1$, works simultaneously in both the three-alcohol mixture and the water system, unambiguously proving the three-component picture of water and the universal validity of our PCA method.

\section{conclusions and discussions}
To conclude, in this work we apply the PCA analysis on X-ray spectrum, and obtain two significant eigenmodes that correspond to three components of water. The dominant first mode corresponds to the conversion between two major components, LDL and HDL, and the second eigenmode corresponds to a third component with five hydrogen bonds and ultra-high density. The 5-H-bond structure makes a separate third component as manifested by its unique non-monotonic response with the external condition, i.e., temperature in our experiment, which is distinct from the monotonic behavior of LDL and HDL. The experimental evidence for the third component makes a breakthrough beyond the prevailing two-component picture, and opens new research directions such as a possible new phase separation for this third component as well as its special properties and interactions with the other two components. The PCA analysis also demonstrates itself as a powerful tool for identifying important components in complex systems.

Similar to the two-component picture, whose proposition brings groundbreaking ideas such as the liquid-liquid phase transition and the second critical point, our experimental discovery of the third component could also open new and exciting research directions, including but not limited to, other possible phase separations related to the third component, unique interactions and rich conversion behaviors among the three components, and the distinct influence of the third component on water properties, such as density, heat capacity, compressibility, and crystallization. Due to the existence of the third component, water's structure becomes more complex and intriguing, and novel theoretical ideas may also naturally arise. As a result, we expect more relevant studies, both in the experimental and theoretical fronts, to appear in the near future.

\section*{Acknowledgements}
The in situ synchrotron XRD was performed at the BL19U2 station of the Shanghai Synchrotron Radiation Facility (SSRF). We thank valuable suggestions from Dr. Na Li, Dr. Guangfeng Liu, Dr. Wentao Li, Dr. Xinhui Lu, Dr. Rui Sun and Dr. Jizhou Li. L. X. acknowledges the financial support GRF14306518, GRF14306920, CRF-C1018-17G, NSFC12074325, CUHK direct grant 4053354; Z. L. acknowledges the financial support from the National Key R $\&$ D program (2016YFA0300102) and the National Nature Science Foundation of China (11675179); Guo Chen acknowledges the financial support from the Natural Science Foundation Project of CQ CSTC (Grant No. cstc2020jcyj-msxmX0106), and the Fundamental Research Funds for the Central Universities (2020CDJ-LHSS-002). Questions should be forwarded to: xuleixu@cuhk.edu.hk, wezer@cqu.edu.cn, or zlluo@ustc.edu.cn.


\begin{figure*}[htpb]
\begin{center}
\centerline{\includegraphics[width=17.2cm]{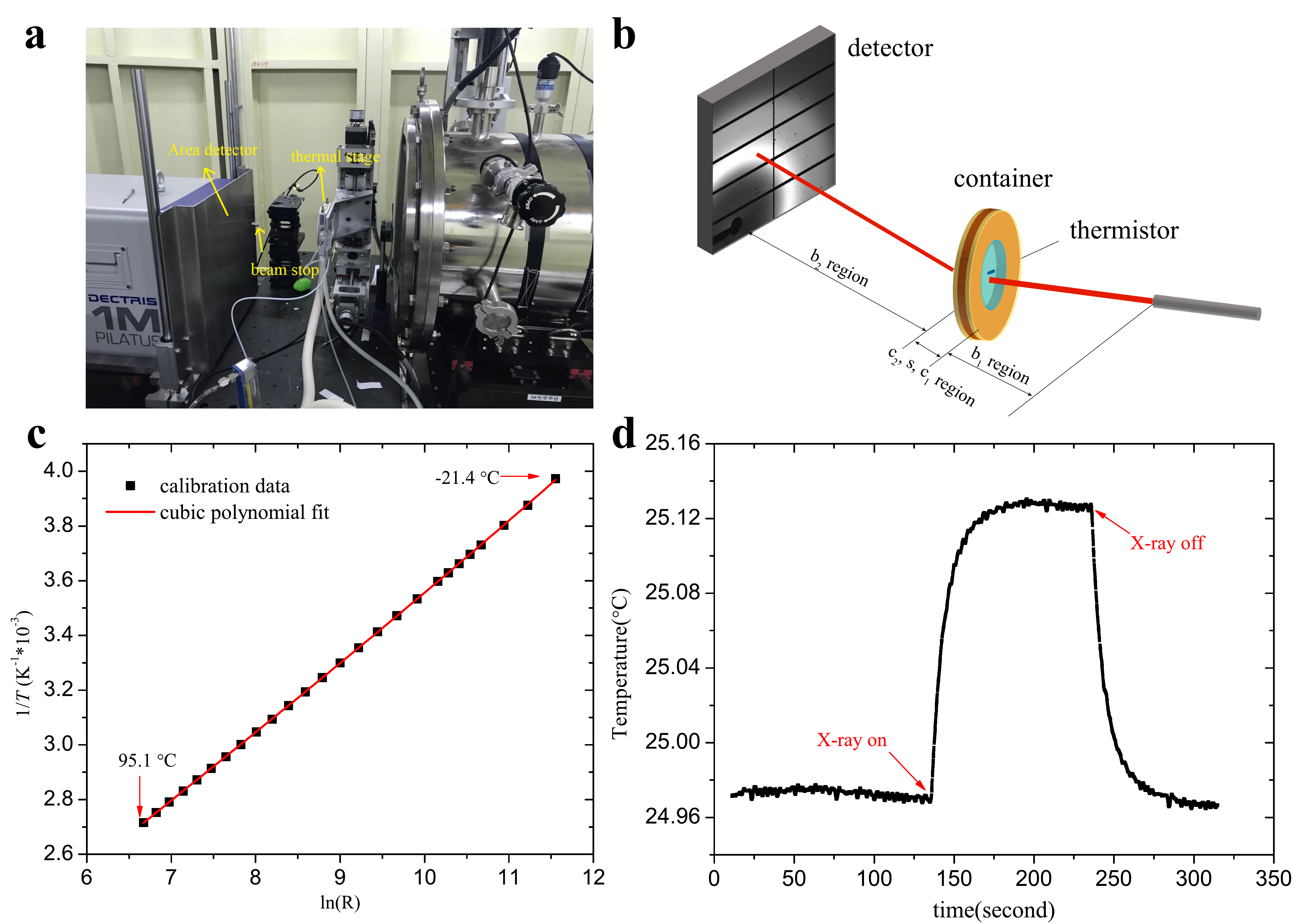}}
\caption{\label{fig:s1}(a) The experimental setup at SSRF. (b) The simplified schematics of experimental setup. The sample was sealed in a copper ring container with a thermistor detecting the real-time temperature. The area detector can collect scattered X-ray from different regions. (c) The calibration curve between $1/T$ and ln($R$) for the temperature probe (thermistor). The curve is excellently fitted by a cubic polynomial between -20 $^\circ$C to 95 $^\circ$C. (d) The real-time temperature of the sample. The laser heating effect is clearly observed indicating the sensitivity of the thermistor.}
\end{center}
\end{figure*}

\section*{APPENDIX A: METHODS}
PCA is a powerful mathematical tool for analyzing data with intrinsic connections. The main idea of PCA is to reduce the data dimensionality by projecting each data point onto only the first few principal components or eigenmodes and obtain lower-dimensional data while preserving as much of the data's variation information as possible. Therefore, the eigenmodes in fact represent the directions towards which the data set vary most substantially. For a system with three different components such as water, the main reason for the system variation is the mutual conversion between the two major components, LDL and HDL. The second reason for the system variation is related to the fraction change in the third component, which has a fraction less than $10\%$. To simplify the question theoretically, we can approximate the mixture's signal as a linear combination of three different components' (see Fig.~5(c)), and the mixture's signal change is due to the fraction variations of these components. In terms of vector expression, the signal of an arbitrary mixture sample $i$ can be expressed as: ${\overrightarrow {I_i}}(q) = {f_{1i}}\overrightarrow \alpha   + {f_{2i}}\overrightarrow \beta   + {f_{3i}}\overrightarrow \gamma  $, with $\overrightarrow \alpha$, $\overrightarrow \beta$, $\overrightarrow \gamma$ the signals of pure components and $f$'s their corresponding fractions. In the water system, LDL and HDL dominate and thus ${f_{1i}} + {f_{2i}} \approx 1$ and ${f_{3i}} \ll 1$. To get the relation between projection pre-factors and component fractions, we can project each sample curve to the eigenmodes, that is $p_{ij} = {\overrightarrow {I_i}}(q)\cdot \overrightarrow {e_{j}} $, where $p_{ij}$ is the projection pre-factor of the ith sample onto the jth eigenmode. According to the previous analysis in Fig.~1(e) inset, the 1st eigenmode can be approximated as $\overrightarrow {(\alpha }  - \overrightarrow {\beta )}$ and thus its unit vector can be approximated as: ${\overrightarrow {e_1}} \approx \overrightarrow {(\alpha }  - \overrightarrow {\beta )} /\left| {\overrightarrow \alpha   - \overrightarrow \beta  } \right|$. Next we derive the linear relationship between $p1$ and $f_1$:
\begin{equation}
\begin{aligned}
{p_{i1}} &= \overrightarrow {{I_i}} (q) \cdot \overrightarrow {{e_1}}  \approx ({f_{1i}}\overrightarrow \alpha   + {f_{2i}}\overrightarrow \beta   + {f_{3i}}\overrightarrow \gamma  ) \cdot \frac{{\overrightarrow \alpha   - \overrightarrow \beta  }}{{\left| {\overrightarrow \alpha   - \overrightarrow \beta  } \right|}}\\
& = \left| {\overrightarrow \alpha   - \overrightarrow \beta  } \right|{f_{1i}} + {f_{3i}}(\overrightarrow \gamma   - \overrightarrow \beta  ) \cdot \frac{{(\overrightarrow \alpha   - \overrightarrow \beta  )}}{{\left| {\overrightarrow \alpha   - \overrightarrow \beta  } \right|}}\\
& + \overrightarrow \beta   \cdot (\overrightarrow \alpha   - \overrightarrow \beta  )\\
& = \left| {\overrightarrow \alpha   - \overrightarrow \beta  } \right|{f_{1i}} + {C_1} + {C_2}
 \end{aligned}
\label{eq:2}
\end{equation}

\noindent The latter two terms do not influence the linear relationship because ${f_{3i}} \ll 1$ and the last term is a constant. This linear relation is confirmed excellently by our experiment and simulation.

We further derive the relationship between $p2$ and $f_3$, by projecting the data curve onto the 2nd eigenmode:
\begin{equation}
\begin{aligned}
{p_{i2}} & = {\overrightarrow {I_i}}(q) \cdot {\overrightarrow {e_2}} = ({f_{1i}}\overrightarrow \alpha   + {f_{2i}}\overrightarrow \beta   + {f_{3i}}\overrightarrow \gamma  ) \cdot {\overrightarrow {e_2}}\\
 & = (\overrightarrow \alpha   - \overrightarrow \beta  ) \cdot {\overrightarrow {e_2}}{f_{1i}} + (\overrightarrow \gamma   - \overrightarrow \beta  ) \cdot {\overrightarrow {e_2}}{f_{3i}}{\rm{ + }}\overrightarrow \beta   \cdot {\overrightarrow {e_2}}\\
& = [\frac{{(\overrightarrow \gamma   - \overrightarrow \beta  ) \cdot {{\overrightarrow {e_2} }}}}{{(\overrightarrow \beta   - \overrightarrow \alpha  ) \cdot {{\overrightarrow {e_2} }}}}{f_{3i}} - {f_{1i}}]C+C'
 \end{aligned}
\label{eq:3}
\end{equation}

\noindent Here $C$ and $C'$ are non-important constants and we only focus on the terms in the square brackets. Because all eigenmodes are orthogonal to each other, i.e., ${\overrightarrow {e_1}} \cdot {\overrightarrow {e_2}} = 0$, and ${\overrightarrow {e_1}} \approx \overrightarrow {(\alpha }  - \overrightarrow {\beta )} /\left| {\overrightarrow \alpha   - \overrightarrow \beta  } \right|$, then the denominator of the pre-factor of $f_{3i}$, $(\overrightarrow \beta   - \overrightarrow \alpha  ) \cdot {\overrightarrow {e_2}}$, is close to zero, resulting in this pre-factor much larger than 1. Thus $p2$'s behavior will strongly depend on $f_3$ due to this large pre-factor, consistent with our experiment and simulation results in Fig.~4(a) and Fig.~5(h) insets. Moreover, this pre-factor is a universal constant in both water system and the three-alcohol system, which gives a universal relation: $p2\propto20.63\times f_3-f_1$ in both systems. This result is excellently confirmed by the main panels of Fig.~4(a) and Fig.~5(h).

To test the robustness of our PCA approach even further, we theoretically construct three-component systems with different Gaussian peaks and their superpositions, and obtain the same results as previous water and three-alcohol systems, as shown later in Appendix M. All these tests unambiguously indicate that the PCA analysis is robust and universal. Therefore, PCA's conclusion that water is a mixture of three components and their fractions vary with temperature are reliable and robust, which should help to settle the long-time debate of water structure in the literature ~\cite{huang2009inhomogeneous,clark2010small}.

\section*{APPENDIX B: Experimental details}
To explore bulk water's structure at different temperatures, we designed a transmission XRD experiment. The setup at BL19U2 station of Shanghai Synchrotron Radiation Facility (SSRF) is shown in Fig.~\ref{fig:s1}(a). The deionized water was sealed in a copper ring (inner radius 8 mm, thickness 1 mm), and two thin Kapton films were used as the windows to make the X-ray get through. A small hole is drilled in the side wall of the copper ring for water injection and temperature detection. The setup (i.e., water inside copper ring) was attached to a thermal stage (Linkam, HFSX350) which can adjust the temperature between -196 $^\circ$C to 350 $^\circ$C. The thermal stage can also provide nitrogen atmosphere to prevent the Kaptpom film window from fogging when the sample is cooled to low temperatures. A thermistor (TE connectivity, Micro-BetaCHIP) was inserted into the water to detect the real-time temperature. The distance between the thermistor and the X-ray area is less than 0.5 mm to make sure that the measured temperature is accurate. An area detector (Pilatus 1M, 169*179 mm$^2$) is used to collect the scattered X-ray. The distance between the detector and the container is calibrated by the silver behenate powder, which is 234.6 mm. The X-ray beam has the spot size 320*43 $\upmu$m$^{2}$ and energy 12.000 $\pm$ 0.002 keV, and the $q$ range collected by the detector is 0.3$\sim$4.5 \AA$^{-1}$. As shown in Fig.~\ref{fig:s1}(b): besides the DI water in the container, the X-ray beam also goes through other materials in other regions, such as air and Kapton film. Therefore, the detector also collects X-ray scattered by these materials and a careful data correction is required for the diffraction raw images. The correction details will be discussed in the next section.

Compared with the temperature set by the thermal stage, the temperature measured by the thermistor inserted into the water is more accurate. The thermistor was calibrated in advance in the circulating bath system (PolyScience, PD15R-40) between -20 $^\circ$C to 95 $^\circ$C. The relationship between $1/T$ and ln($R$) ($R$ is thermistor's resistance) can be fitted with a cubic polynomial (show in Fig.~\ref{fig:s1}(c)). We used a mutimeter (Keithley, model 2700) to measure the real-time resistance of the thermistor with a frequency up to 30 Hz, and the temperature can be obtained with the calibration curve Fig.~\ref{fig:s1}(c). A typical temperature measurement curve is shown in Fig.~\ref{fig:s1}(d): there is a slight heating effect below 0.2 $^\circ$C after the X-ray is turned on. Apparently, we can monitor the sample temperature in real time with a high accuracy below 0.1 $^\circ$C.

\section*{APPENDIX C: Experiment data correction}
When water was measured in the copper-ring container, as shown in Fig.~\ref{fig:s1}(b), the scattering intensity collected can be represented as the following (multiple scattering was neglected):
\begin{equation}
\begin{array}{lll}
{I_{scb}}(\theta ) = {I_{b1}}(\theta )\\
+ {I_{c1}}(\theta )\cdot att_c(\theta ,{\mu _c},{t_{c1}})\cdot {e^{ - ({\mu _s}{t_s} + {\mu _c}{t_{c2}} + {\mu _b}{t_{b2}})/cos\theta }}\\
 + {I_s}(\theta )\cdot {T_{c1}}\cdot att_s(\theta ,{\mu _s},{t_s})\cdot {e^{ - ({\mu _c}{t_{c2}} + {\mu _b}{t_{b2}})/cos\theta }} \\
 + {I_{c2}}(\theta )\cdot {T_{c1}}\cdot {T_s}\cdot att_c(\theta ,{\mu _c},{t_{c2}})\cdot {e^{ - {\mu _b}{t_{b2}}/cos\theta }}\\
 + {I_{b2}}(\theta )\cdot {T_{c1}}\cdot {T_s}\cdot {T_{c2}}\cdot att_b(\theta ,{\mu _b},{t_{b2}})
\label{S1}
\end{array}
\end{equation}
where the subscripts $s$, $c$ and $b$ represent contributions from sample (i.e., water), container windows (i.e., Kapton film) and background air respectively. 1 and 2 represent different regions along the light path (see Fig.~\ref{fig:s1}(b)). $\mu$ is the attenuation coefficient. $t$ is the thickness of the corresponding material. $\theta$ is the scattering angle. $att(\theta ,\mu ,t)$ is the angle dependence of the attenuation when the scattering X-ray goes through the corresponding material. $att(\theta ,\mu ,t)$ in general also depends on the incident X-ray energy, but in our experiment the energy is fixed. $T$ is the $\theta$ = 0 transmission factor, typically represented as ${e^{ - \mu t}}$. The final exponential part in relevant terms represents the absorption by the following up materials after the current scattering material: the scattered light needs to go through these following up materials and gets absorbed before reaching the detector. When there was no water in the container and the container space (labelled by $b3$) was filled with air, the corresponding background scattering intensity is:
\begin{equation}
\begin{array}{lll}
{I_{cb}}(\theta ) = {I_{b1}}(\theta )\\
+ {I_{c1}}(\theta )\cdot att_c(\theta ,{\mu _c},{t_{c1}})\cdot {e^{ - ({\mu _b}{t_s} + {\mu _c}{t_{c2}} + {\mu _b}{t_{b2}})/cos\theta }}\\
 + {I_{b3}}(\theta )\cdot {T_{c1}}\cdot att_b(\theta ,{\mu _b},{t_s})\cdot {e^{ - ({\mu _c}{t_{c2}} + {\mu _b}{t_{b2}})/cos\theta }} \\
 + {I_{c2}}(\theta )\cdot {T_{c1}}\cdot {T_{b3}}\cdot att_c(\theta ,{\mu _c},{t_{c2}})\cdot {e^{ - {\mu _b}{t_{b2}}/cos\theta }}\\
 + {I_{b2}}(\theta )\cdot {T_{c1}}\cdot {T_{b3}}\cdot {T_{c2}}\cdot att_b(\theta ,{\mu _b},{t_{b2}})
 \label{S2}
\end{array}
\end{equation}
In our experiment, the $I_{b1}$ term was prevented from entering the detector and thus can be neglected. The attenuation coefficient for sample water, container windows (Kapton film) and background air at 20 $^\circ$C are 2.9079 cm$^{-1}$, 2.454 cm$^{-1}$ and 3.349*10$^{-3}$ cm$^{-1}$ respectively. Their thicknesses are 0.1 cm, 0.005 cm and 23.46 cm respectively. We can find that ${\mu _s}{t_s} \gg {\mu _c}{t_c},{\mu _s}{t_s} \gg {\mu _b}{t_b}$, after combining Eq.~(\ref{S1}) and Eq.~(\ref{S2}), the intensity contributed by the sample can be calculated as~\cite{skinner2012area}:
\begin{equation}
\begin{array}{lll}
{I_s}(\theta ) \approx \frac{{{I_{scb}}(\theta ) - {T_s}\cdot {I_{cb}}(\theta )}}{{att_s(\theta ,{\mu _s},{t_s})}}
\label{S3}
\end{array}
\end{equation}
Because all the materials the X-ray goes through are slab shaped with uniform thickness, the attenuation for this geometry is:
\begin{equation}
\begin{array}{lll}
att(\theta ,\mu ,t) = \frac{{\exp (\frac{{ - \mu t}}{{\sin (\theta  + \alpha )}})\left[ {\exp (\frac{{\mu t}}{{\sin (\theta  + \alpha )}} - \frac{{\mu t}}{{\sin \alpha }}) - 1} \right]}}{{\frac{{\mu t}}{{\sin (\theta  + \alpha )}} - \frac{{\mu t}}{{\sin \alpha }}}}
\label{S4}
\end{array}
\end{equation}
where $\alpha$ is the angle between the incident light and the slab surface, which is $\pi$/2 in our experiment, and thus Eq.~({\ref{S4}}) simplifies to:
\begin{equation}
\begin{array}{lll}
att(\theta ,\mu ,t) = \frac{{\exp ( - \mu t) - \exp( - \mu t/\cos \theta )}}{{\mu t(1/cos\theta  - 1)}}
\label{S5}
\end{array}
\end{equation}
Because of the geometry of the area detector and the polarization of the X-ray beam, the geometric and polarization~\cite{kahn1982macromolecular} corrections are required:
\begin{equation}
\begin{array}{lll}
I_s^c(\theta ) = \frac{{{I_s}(\theta )geo(\theta )}}{{pol(\theta )}}\\
geo(\theta ) = co{s^3}(\theta )\\
pol(\theta ) = \frac{1}{2}(1 + {\cos ^2}\theta  - f\cos \varphi {\sin ^2}\theta )
\label{S6}
\end{array}
\end{equation}
where $\theta$ is the scattering angle, $\varphi$ is the azimuthal angle, and $f$ is the polarization factor of the X-ray beam, which is 0.99 in our experiment. The geometric and polarization corrections were performed by the software Fit2D~\cite{hammersley2016fit2d}, which was used to integrate the X-ray diffraction images into $I(q)$ curves.

\begin{figure}[htpb]
\begin{center}
\centerline{\includegraphics[width=8.6cm]{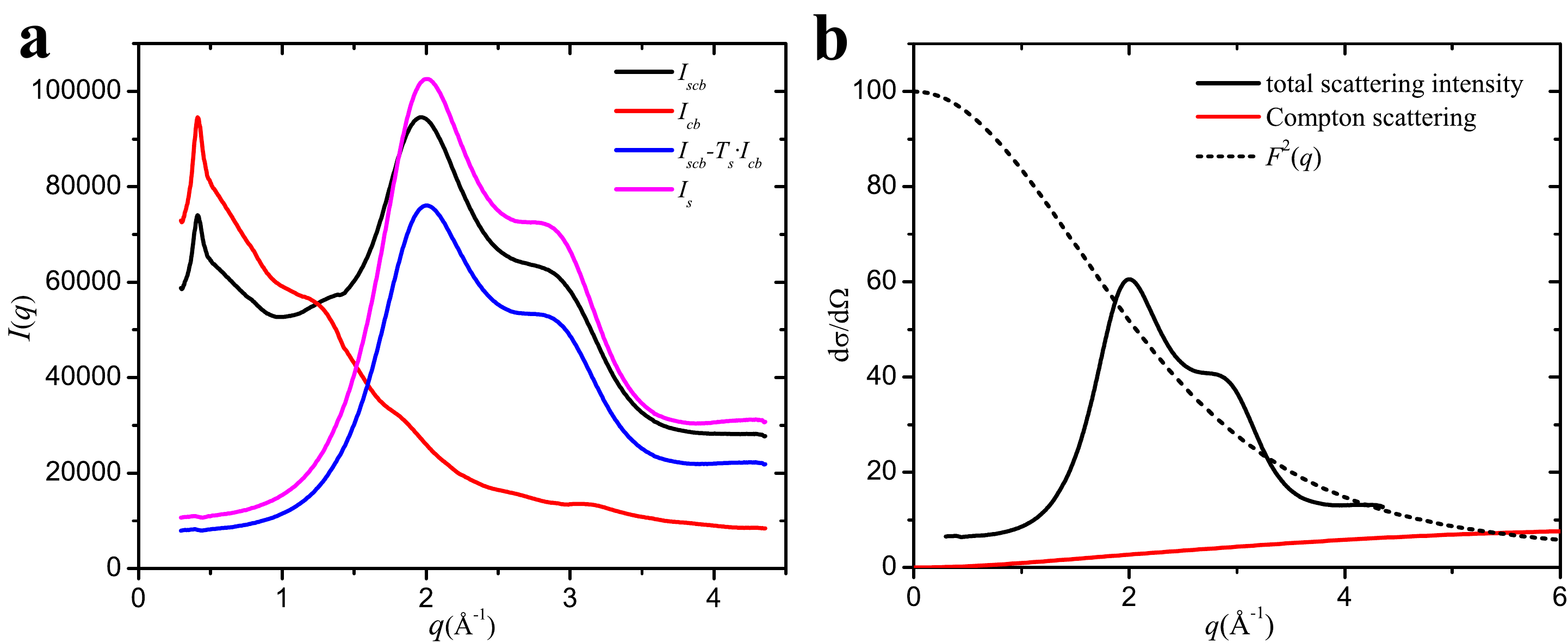}}
\caption{\label{fig:s2}(a) The integrated intensity before and after data correction: the original raw data without corrections ($I_{scb}$, black), intensity from the background air and sample container ($I_{cb}$, red), intensity after subtracting the background ($I_{scb}-T_s\cdot I_{cb}$, blue), the final intensity after slab absorption correction ($I_s$, pink). (b) The necessary data needed to calculate the molecular structure factor $S(q)$. The fully corrected scattering intensity at 25 $^\circ$C (black solid). The square of water molecular form factor calculated by quantum mechanics~\cite{wang1994chemical} (black dash). Compton scattering in theory (red).}
\end{center}
\end{figure}

However, Eq.~(\ref{S3}) cannot be directly used to subtract the background, because the incident light intensity is always fluctuating with time and a correction factor based on the actual intensity is required. To obtain this correction factor, we notice that there is a sharp peak around $q$ = 0.4 \AA$^{-1}$ in the intensity curve $I_{scb}$ (see Fig.~\ref{fig:s2}(a)), which comes from the Kapton film instead of liquid water. Therefore, we multiply a correction factor $T_s$ to make sure that after background subtraction, the true intensity curve around $q$ = 0.4 \AA$^{-1}$ is completely flat without any peak feature~\cite{hura2003water}. The profiles of the intensity curves before and after corrections are presented in Fig.~\ref{fig:s2}(b).

The fully corrected intensity can be normalized by the number of water molecules, the incident beam flux and the exposure time. In practice the KroghMoe~\cite{krogh1956method} and Norman~\cite{norman1957fourier} methods are generally used because of convenience. The scale factor $\alpha$ is defined as:
\begin{equation}
\begin{array}{lll}
\alpha {\rm{ = }}\frac{{{\rm{ - }}2{\pi ^2}{z^2}\rho {\rm{ + }}\int\limits_0^{{q_{\max }}} {\left[ {{F^2}(q) + {I_{incoherent}}(q)} \right]{q^2}{\rm{d}}q} }}{{\int\limits_0^{{q_{\max }}} {{I_{experiment}}(q){q^2}{\rm{d}}q} }}
\label{S7}
\end{array}
\end{equation}
where $\rho$ is the number density of molecules (in molecules/\AA$^{3}$), $z$ is the number of electrons per water molecule, $F^{2}(q)$ is the square of water molecule's form factor, $q_{max}$ is the maximum wavenumber we can measure in the experiment, $I_{incoherent}(q)$ is the incoherent or Compton scattering from the sample, and $I_{experiment}(q)$ is the fully corrected intensity measured from our experiment. Both $F^{2}(q)$ and $I_{incoherent}(q)$ can be calculated theoretically by quantum mechanics~\cite{wang1994chemical}. In our experiment, the minimum $q$ is 0.3 \AA$^{-1}$ because of the beam stop, and thus the lower limit of the integration is 0.3 \AA$^{-1}$ instead of zero. After normalized by $\alpha$, the intensity is re-scaled onto a universal scale in the electron units, and then the Compton scattering~\cite{wang1994chemical} is subtracted. One typical example of the normalized intensity after subtracting Compton scattering is illustrated in Fig.~\ref{fig:s2}(b) by the pink $I_s$ curve. Using $S(q) - 1 = [I^{nor}(q) - {F^2}(q)]/[2{f_H}(q) + {f_O}(q)]^2$ , the molecular structure factor $S(q)$ - 1 at different temperatures can be calculated~\cite{skinner2013benchmark}. Following the work in the previous research~\cite{skinner2012area}, we start from the independent atom approximation (IAA) form factors of hydrogen and oxygen atoms using the 6-Gaussian fitting functions~\cite{su1997relativistic}. After considering the electron density change caused by the chemical bonding, the modified atomic form factors (MAFF), ${f_\alpha }(q) = f_\alpha ^0(q)\left[ {1 - (a_\alpha/z_\alpha)\cdot \exp ( - {q^2}/2{\delta ^2})} \right]$, are then calculated, where $f_\alpha ^0(q)$ is the IAA form factor, $z_{\alpha}$ is the atomic number for atom $\alpha$, and $a_{\alpha}$ represents the electron transfer, which is -1 for oxygen atom and +0.5 for hydrogen atom according to the previous literature~\cite{skinner2013benchmark}. The value of $\delta$ is 2.0 \AA$^{-1}$ for both oxygen and hydrogen atoms.

\section*{APPENDIX D: PCA analysis}
The PCA analysis is done on the structure factor curves $S(q)$-1 at different temperatures. We discretize every $S(q)$-1 curve into 915 discrete values with equal partition in our experimental range between 0.3 to 4.5 \AA$^{-1}$. Thus each $S(q)$-1 curve at one specific temperature is converted into a 1(row)*915(column) vector. We measured 18 different temperatures between -5 $^\circ$C to 80 $^\circ$C in our experiment, and all the 18 curves are converted to a 18*915 data matrix. Then this data matrix's covariance matrix (915*915 in dimension) is constructed and the covariance matrix's eigenmodes and eigenvalues are calculated~\cite{henkes2012extracting,chen2010low}, as shown in Fig. 1(e) and (f) in the main text. These calculations are performed by the professional mathematical software Matlab~\cite{matlabpca}. Note that the partition number (915 in our case) chosen to discretize $S(q)-1$ curves has no significant influence on the final PCA results, as long as the partition number is large enough to reflect the curve's true profile. We also tried other partition numbers and the PCA results are the same.

\begin{figure}[htpb]
\begin{center}
\centerline{\includegraphics[width=8.6cm]{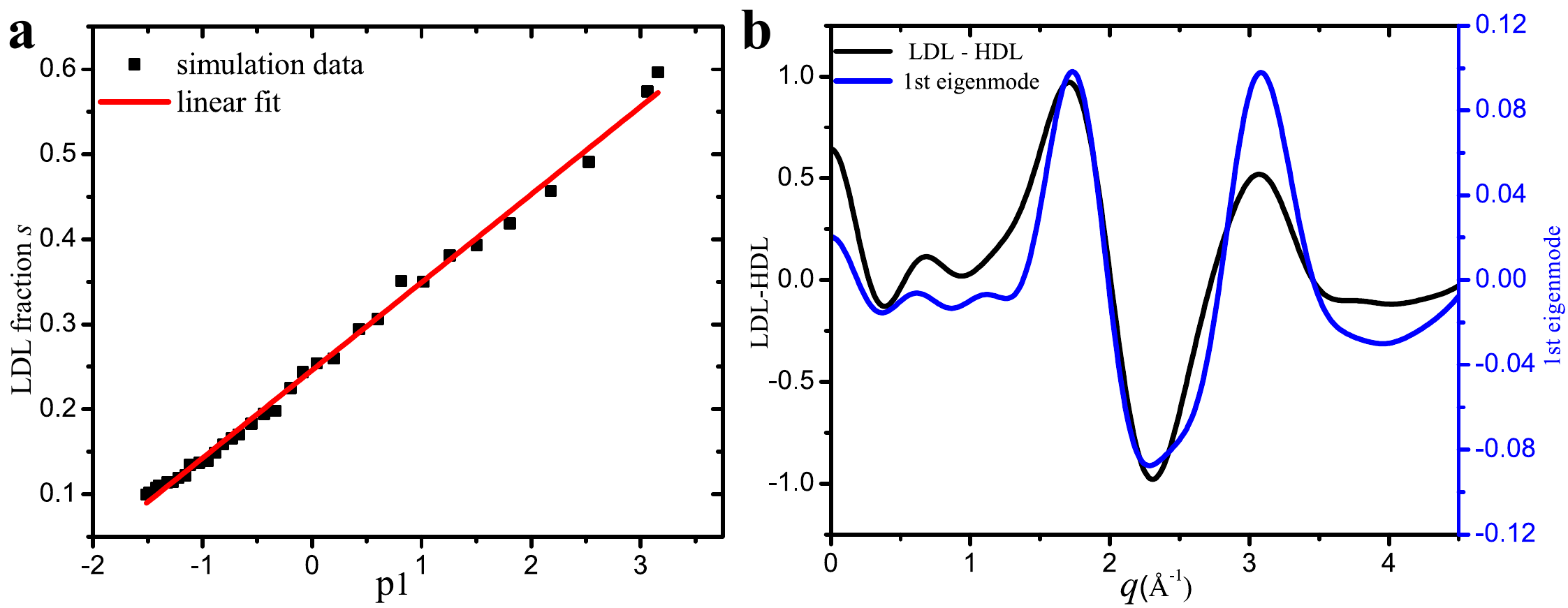}}
\caption{\label{fig:s3}PCA analysis on our numerical simulations, which covers a much broader temperature range (-60 $^\circ$C to 100 $^\circ$C) than the experiment (-5 $^\circ$C to 80 $^\circ$C). (a) The LDL fraction shows an excellent linear relation with the projection pre-factor $p1$. (b) The first eigenmode agrees very well with the curve LDL minus HDL.}
\end{center}
\end{figure}

In addition, more data sets across a broader temperature range can increase the accuracy of PCA results. Therefore, we also performed the PCA analysis on the structure factor curves from numerical simulations, which cover a much broader temperature range between -60 $^\circ$C and 100 $^\circ$C with 33 temperatures in between. Consistent with the experimental results, the fraction of LDL, $s$, has a good linear relationship with $p1$, and the 1st eigenmode has very similar profile as the structure factor difference between LDL and HDL, as shown in Fig.~\ref{fig:s3}.

\section*{APPENDIX E: Simulation details}
Many different water models exist. According to the number of interaction points, there are 3-site models SPC~\cite{berendsen1981intermolecular}, TIP3P~\cite{jorgensen1983comparison}, SPC/E~\cite{berendsen1987missing}, 4-site models TIP4P~\cite{jorgensen1983comparison}, TIP4P-Ew~\cite{horn2004development}, TIP4P/2005~\cite{abascal2005general}, 5-site models ST2~\cite{stillinger1974improved}, and TIP5P~\cite{mahoney2000five}, etc.. In general, the electrostatic interaction is calculated with Coulomb's law, and the dispersion and repulsion forces are calculated with the Lennard-Jones potential~\cite{allen2017computer}. The main differences between these models are the charges carried by the sites, the bond lengths, and the parameters in the Lennard-Jones potential besides the number of the sites. We tried five models, SPC/E, TIP3P, TIP4P, TIP4P-Ew and TIP5P in our MD simulations, with the software Gromacs (v.2016.5)~\cite{berendsen1995gromacs}. Consistent with experimental conditions, the simulations were performed under the isothermal-isobaric NPT ensemble. The isotropic Parrinello-Rahman barostat~\cite{nose1983constant,parrinello1981polymorphic} and the Nose-Hoover thermostat~\cite{hoover1985canonical,nose1984molecular} were used to guarantee the simulations were performed at the set pressure and temperature. 512 water molecules were put in a cubic box (side length $\approx$~25~\AA) and the pressure was kept fixed at the atmospheric pressure 1.01 bar and the temperature varied from -60 $^\circ$C to 100 $^\circ$C in every five degrees. Periodical boundary condition (PBC) was used to eliminate the influence from boundary. The simulation time for water at low (below -20 $^\circ$C) and high (above -20 $^\circ$C) temperatures last 20 and 10 ns respectively, which are long enough for the system to reach the equilibrium state. Besides the normal pressure, the simulation time for water at extreme HDL and LDL conditions (super-high and super-low pressure) was 20 ns.

\begin{figure}[htpb]
\begin{center}
\centerline{\includegraphics[width=8.6cm]{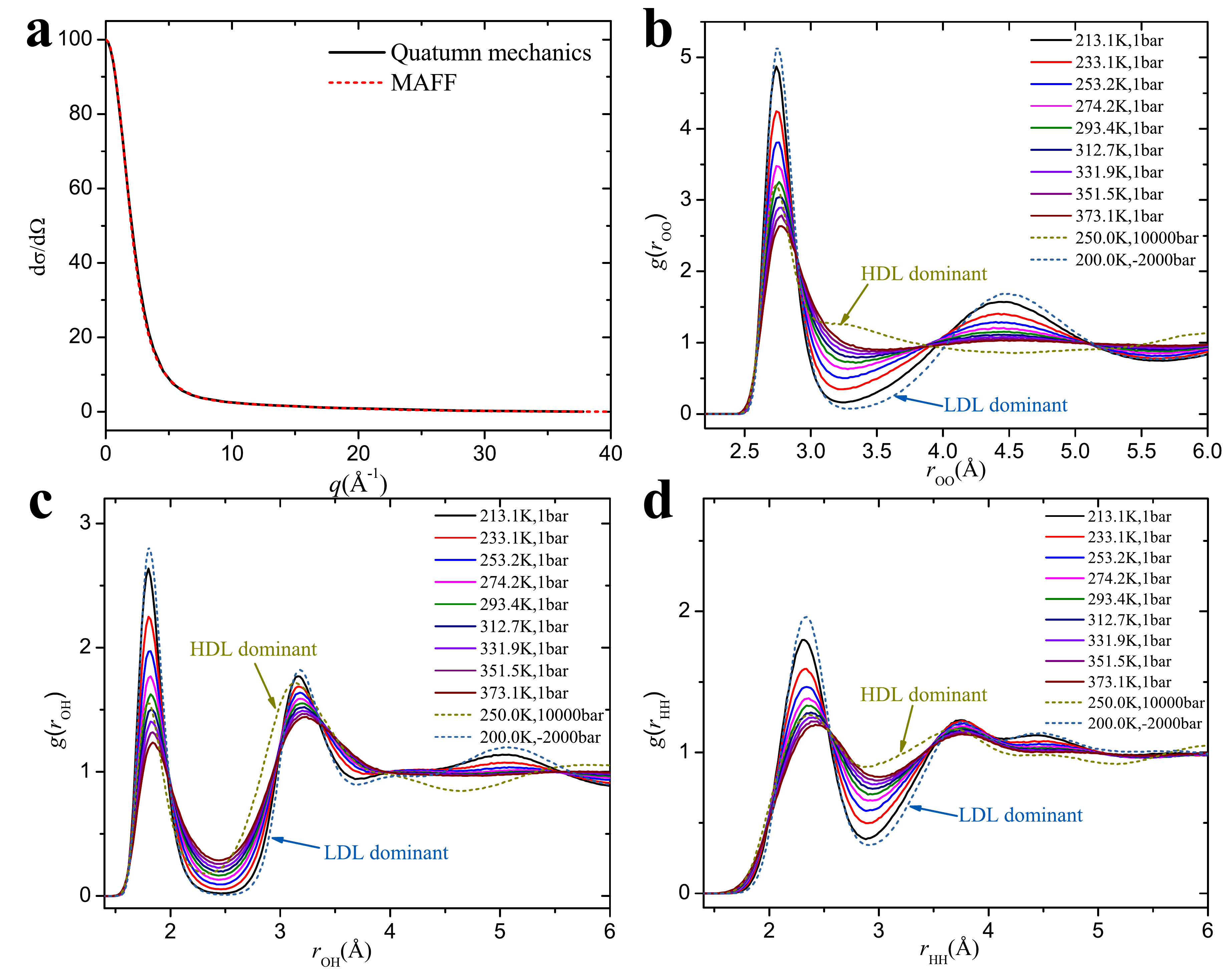}}
\caption{\label{fig:s4}(a) The $F^{2}(q)$ calculated using quantum mechanics and MAFF agree very well within a broad $q$ range. (b), (c) and (d) The related radial distribution functions, $g(r_{OO})$, $g(r_{OH})$ and $g(r_{HH})$, which are required to calculate the simulated scattering intensity at different conditions.}
\end{center}
\end{figure}

\section*{APPENDIX F: Simulated scattering intensity and structure factor}
In general, the total X-ray differential scattering cross section d$\sigma$/d$\Omega$ of molecular liquids can be divided into two parts, contributions from individual molecules (self-scattering) and from intermolecular correlations defined by the function $H_{ij}(q)$,
\begin{equation}
\begin{array}{lll}
I(q) = \sum\limits_{ij} {{x_i}{x_j}{f_i}(q){f_j}(q)} \frac{{\sin (q{r_{ij}})}}{{q{r_{ij}}}} \\
+ \sum\limits_{i \le j} {{x_i}{x_j}{f_i}(q){f_j}(q){H_{ij}}(q)}
\label{S8}
\end{array}
\end{equation}
\begin{equation}
\begin{array}{lll}
{H_{ij}}(q) = 4\pi \rho \int\limits_0^\infty  {{r^2}dr({g_{ij}}(r) - 1)} \frac{{\sin (qr)}}{{qr}}
\label{S9}
\end{array}
\end{equation}
where $x_{i}$ is the atomic fraction of atom type $i$, $f_{i}(q)$ is the atomic scattering factor for atom type $i$ and $r_{ij}$ are the intramolecular distances between atom centers. $\rho$ is the atomic density and $g_{ij}(r)$ is the radial distribution function (RDF) between atom $i$ and $j$. In our experiment, the scattering intensity $I(q)$ can be represented as,
\begin{equation}
\begin{array}{lll}
I(q) = I_{self - intra}^{nor}(q) + I_{inter}^{nor}(q)\\
 = f_O^2(q) + 2f_H^2(q) + 4{f_O}(q){f_{\rm{H}}}(q)\frac{{\sin (q{r_{^{OH}}})}}{{q{r_{^{OH}}}}} \\
 + 2f_H^2(q)\frac{{\sin (q{r_{^{HH}}})}}{{q{r_{^{HH}}}}} + \frac{4}{3}H_{HH}^{inter}(q)f_H^2(q) \\
 + \frac{4}{3}H_{OH}^{inter}(q){f_H}(q){f_O}(q) + \frac{1}{3}H_{OO}^{inter}(q)f_O^2(q)
\label{S10}
\end{array}
\end{equation}
The first term in Eq.~(\ref{S10}) is $F^{2}(q)$ mentioned in {\it Experiment data correction} section, which can be calculated using quantum mechanics~\cite{wang1994chemical}. Of course, it can also be calculated using the formula in Eq.~(\ref{S10}). In fact, the difference between the two methods can be neglected when appropriate parameters are used in Eq.~(\ref{S10}), as shown in Fig.~\ref{fig:s4}(a). The bond lengths $r_{OH}$ = 0.9572 \AA~and $r_{HH}$ = 1.5139 \AA~are used. Our simulated results are calculated using the quantum mechanics data. The second term in Eq.~(\ref{S10}) (or Eq.~(\ref{S8})) represents the intermolecular correlations: $H_{ij}$ term can be calculated using Eq.~(\ref{S9}), the atomic density is 512*3 divided by volume, and the upper limit of integral is set to 10 \AA. The related $g_{ij}(r)$ (shown in Fig.~\ref{fig:s4}(b), (c) and (d)) can be calculated after the system reaches equilibrium, and then the scattering intensity can be calculated using Eq.~(\ref{S9}) and~(\ref{S10}). The corresponding molecular structure factor $S(q)$-1 based on $I(q)$ can then be obtained.

To calculate the molecular structure factor of the 5-hydrogen-bond component, $S_{5H}(q)$-1, we first pick out the molecules surrounded by five hydrogen bonds, and set them as origin to obtain $g_{ij}(r)$. Then the intensity can be calculated using Eq.~(\ref{S9}) and (\ref{S10}), which further gives $S_{5H}(q)$-1.

Among all the five popular water models tested, we found that the results from the TIP4P-Ew model agree with our experiments the best (shown in the main text Fig.~2(e) and (f)), which is used throughout our simulation.

\begin{figure}[htpb]
\begin{center}
\centerline{\includegraphics[width=8.6cm]{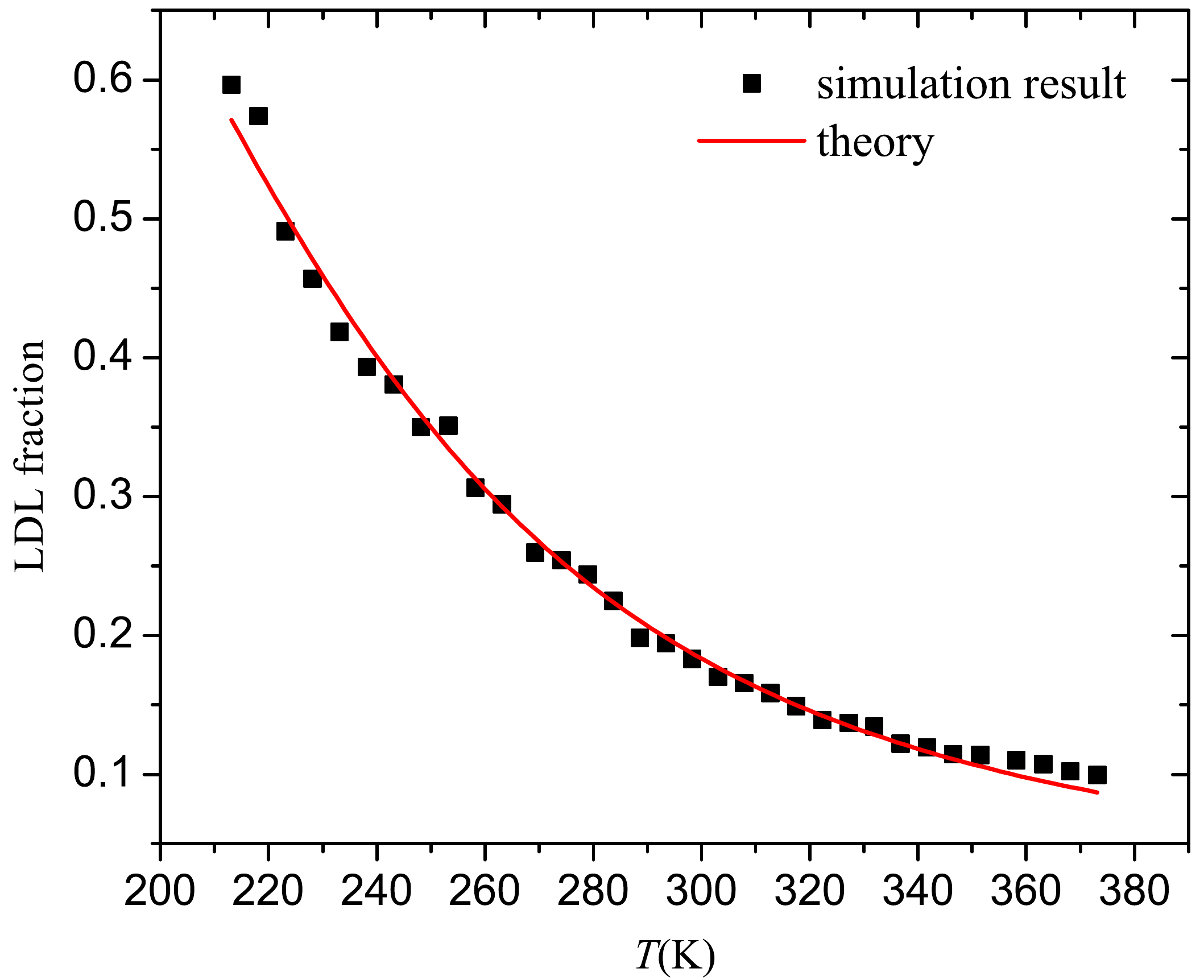}}
\caption{\label{fig:s5}Verifying the two-component model with simulation. The simulation results (square symbols) agree well with the two-component model (solid curve), in a temperature range much broader than the experiment.}
\end{center}
\end{figure}

\section*{APPENDIX G: Two-component model}
For a liquid composed by two components, such as LDL and HDL, its free energy $G$ can be represented as~\cite{tanaka2000thermodynamic}:
\begin{equation}
\begin{array}{lll}
G = {G_{LDL}} + s\Delta G + {k_B}T\left[ {s\ln s + (1 - s)\ln (1 - s)} \right] \\
+ Js(1 - s)
\label{S11}
\end{array}
\end{equation}
where $s$ is the fraction of LDL, $G_{LDL}$ is the free energy of pure LDL, $\Delta G = {G_{LDL}} - {G_{HDL}}{\rm{ = }}\Delta E - T\Delta \sigma  + P\Delta V$, and $J$ is the coupling between the two components. As the two components convert into each other, the equilibrium will be reached under the following condition:
\begin{equation}
\begin{array}{lll}
\frac{{\partial G}}{{\partial s}} = \Delta G + {k_B}T\ln (\frac{s}{{1 - s}}) + J(1 - 2s) = 0
\label{S12}
\end{array}
\end{equation}
Under our experiment and simulation conditions, the terms $J$ and $P\Delta V$ in $\Delta G$ can be neglected~\cite{shi2018common,shi2018origin}, and then the relation between $s$ and $T$ can be obtained:
\begin{equation}
\begin{array}{lll}
s = \frac{1}{{1 + {e^{\frac{{\Delta E}}{{{k_B}T}} - \frac{{\Delta \sigma }}{{{k_B}}}}}}}
\label{S13}
\end{array}
\end{equation}
In fact, this relationship of Eq.~(\ref{S13}) is applicable in a temperature range much broader than our experiment. To verify it, we apply it to the simulations that can reach down to -60 $^\circ$C, and find that the two-component model agrees excellently with the simulations, as shown in Fig.~\ref{fig:s5}. The fitting gives that $\Delta E/k_B$ = -1311.6 K and $\Delta\sigma/k_B$ = -5.87, which are close to our main text results.

\section*{APPENDIX H: Calculating the fractions of LDL and HDL}
\begin{figure}[htpb]
\begin{center}
\centerline{\includegraphics[width=8.6cm]{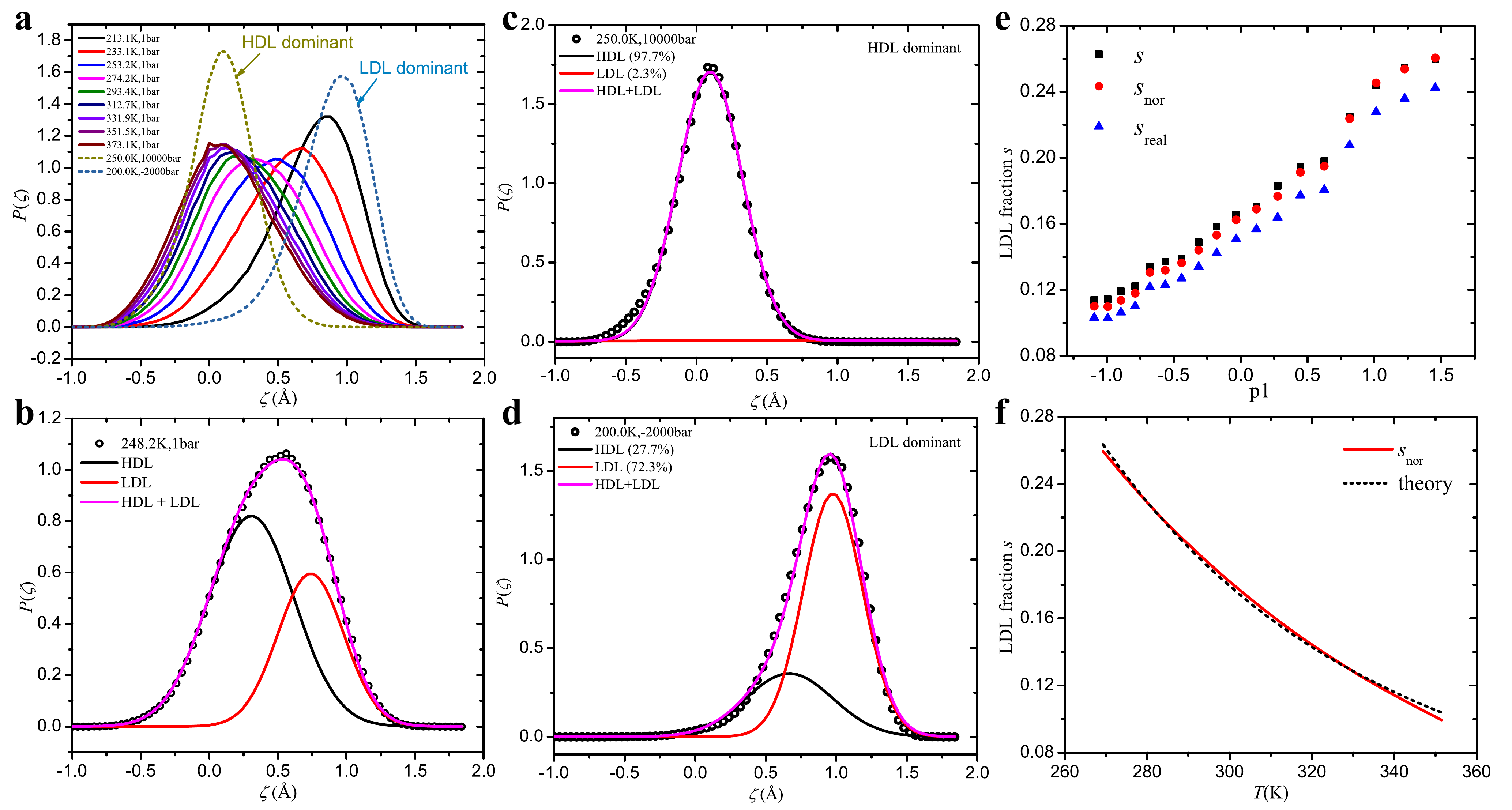}}
\caption{\small\label{fig:s6}(a) The distribution of the order parameter, $P(\zeta)$, under different conditions. (b) Fitting $P(\zeta)$ excellently with two Gaussian curves. The left Gaussian corresponds to HDL and the right one corresponds to LDL. Their weights give the fractions of HDL and LDL. The two-Gaussian fitting method also works well after the 3rd component is considered and excluded. (c) and (d) The two-Gaussian decomposing result of our HDL-dominant and LDL-dominant systems. The result shows that we have obtained a HDL dominant (97.7$\%$) system (c) and an LDL dominant (72.3$\%$) system (d). (e) The fraction of LDL obtained in three different methods: (1) the previous two-component method without considering the 3rd component (black), (2) after considering and excluding the 3rd component and then renormalizing the LDL fraction among the rest (red), and (3) the real LDL fraction in the system after considering the 3rd component (blue). The result shows that the fraction of LDL changes slightly (a few percent) due to the low fraction of the 3rd component (blue). (f) Fitting the re-normalized LDL result with the two-component model, Eq.~(\ref{S13}). The red solid curve comes from the data points of the re-normalized method (red) in (e) and the agreement with theory (dashed curve) is very well.}
\end{center}
\end{figure}

Following the method in ref~\cite{russo2014understanding}, we decompose the order parameter distribution $P(\zeta)$ into the combination of two Gaussian functions,
\begin{equation}
\begin{array}{l}
P(\zeta ) = (1 - s){P_{{\rm{HDL}}}}(\zeta ) + s{P_{{\rm{LDL}}}}(\zeta )\\
 = \frac{{P(0)}}{{\exp ( - \frac{{m_{{\rm{HDL}}}^2}}{{2\sigma _{{\rm{HDL}}}^2}})}}\exp ( - \frac{{{{(\zeta  - {m_{{\rm{HDL}}}})}^2}}}{{2\sigma _{{\rm{HDL}}}^2}}) \\
 + (1 - \frac{{{\sigma _{{\rm{HDL}}}}\sqrt {2\pi } P(0)}}{{\exp ( - \frac{{m_{{\rm{HDL}}}^2}}{{2\sigma _{{\rm{HDL}}}^2}})}})\frac{{\exp( - \frac{{{{(\zeta  - {m_{{\rm{LDL}}}})}^2}}}{{2\sigma _{{\rm{LDL}}}^2}})}}{{\sqrt {2\pi } {\sigma _{{\rm{LDL}}}}}}
\label{S14}
\end{array}
\end{equation}
where $s$ is the fraction of LDL, and this formula has assumed that $P_{LDL}$(0) = 0, which is reasonable because of the open structures of LDL. The distribution of order parameter, $P(\zeta)$, under different temperatures are shown in Fig.~\ref{fig:s6}(a). For comparison, the HDL and LDL dominant samples are also plotted. As temperature increases, the data curve shifts from LDL-like to HDL-like, as a result of their mutual conversion. We can further obtain the exact fractions of LDL and HDL in each sample, by fitting its $P(\zeta)$ curve with two Gaussian curves, as shown in Fig.~\ref{fig:s6}(b): the two Gaussian curves represent thermally broadened HDL and LDL components respectively, and the weight of each Gaussian gives the fraction of each component. Using this formula to fit HDL and LDL dominant systems' $P(\zeta)$, we can get their fractions as shown in Fig.~\ref{fig:s6}(c) and (d).

Due to the existence of the 3rd component, the fractions of LDL and HDL need to be recalculated. We use a similar method as the two-component model, except that we identify the 3rd component and exclude such molecules first. After eliminating the third component and renormalizing the rest into unity, we obtained the renormalized order parameter distribution $P(\zeta)$ and decomposed it into two Gaussian curves to get the new renormalized fractions of LDL and HDL at different temperatures. One typical fitting with two Gaussian curves is shown in Fig.~\ref{fig:s6}(b) and the agreement is excellent. Due to the low fraction of the 3rd component, the new renormalized fractions of LDL (and HDL) only change very little in comparison to the two-component model, as shown by the red disc and black square symbols in Fig.~\ref{fig:s6}(e). The real fractions of LDL after considering the existence of the 3rd component decreases by a few percent in comparison to the previous pure two-component result, which are shown as the blue triangles in Fig.~\ref{fig:s6}(e). Fitting the theory of Eq.~(\ref{S13}) with the renormalized $s$ curve in Fig.~\ref{fig:s6}(f), we can obtain the fitting parameters of $\Delta E/k_B$ = -1296.6 K and $\Delta\sigma/k_B$ = -5.84, which only deviate slightly (around 5$\%$) from the previous two-component model ($\Delta E/k_B$ = -1238.3 K, $\Delta\sigma/k_B$ = -5.63). Therefore, the third component does not cause significant changes and the two-component model still describes the main feature of water's structure.

\begin{figure}[htpb]
\begin{center}
\centerline{\includegraphics[width=8.6cm]{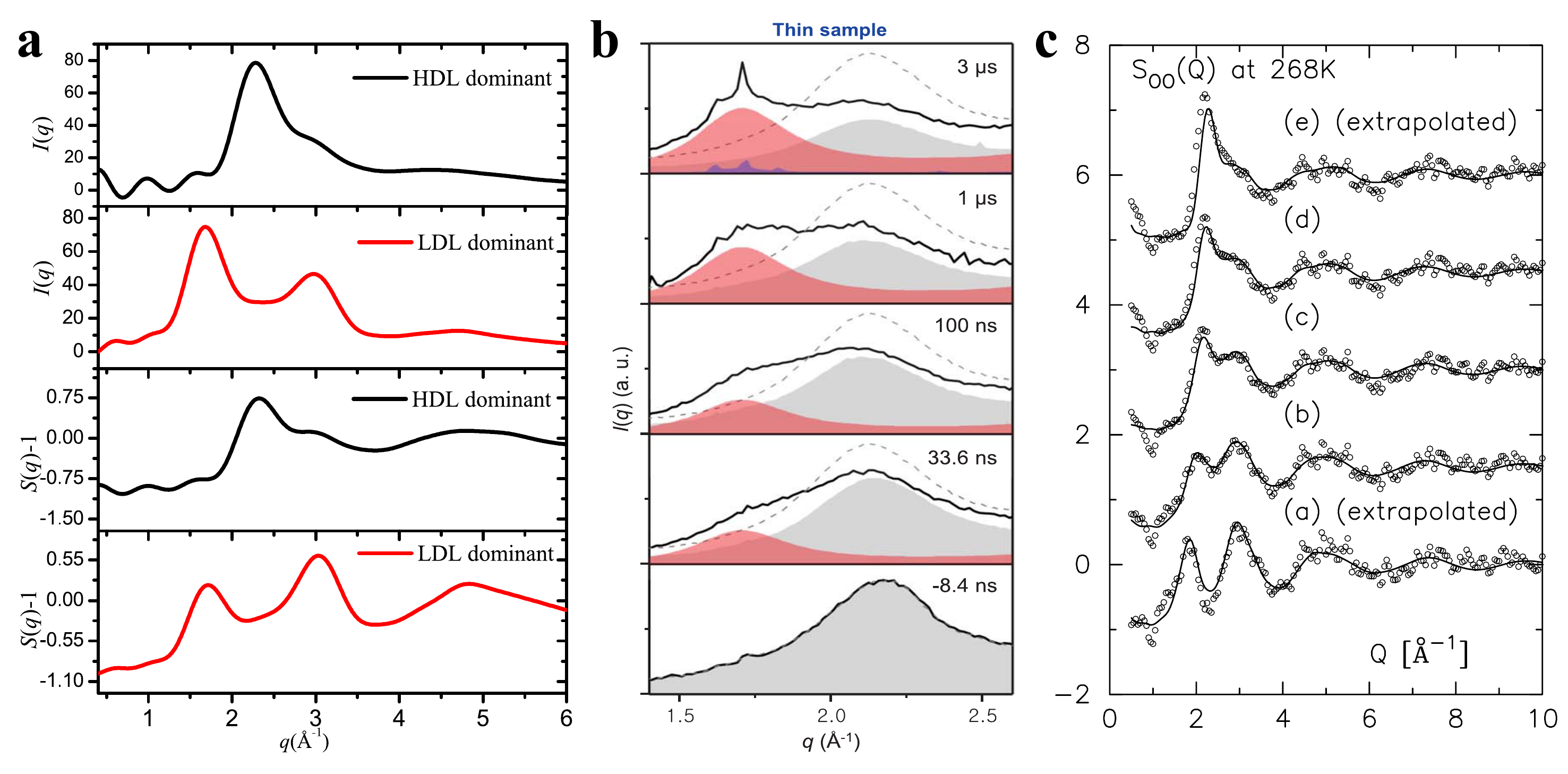}}
\caption{\label{fig:s7} Comparison between our simulated LDL and HDL system in (a) with the previous experiment data in (b)~\cite{kim2020experimental} and (c)~\cite{soper2000structures}. (a) Our simulated scattering intensity ($I(q)$) and molecular structure factor ($S(q)$) of HDL-dominant and LDL-dominant systems. (b) The contribution of HDL and LDL to the scattering intensity in ref~\cite{kim2020experimental}. The LDL curve (red) shows a peak at $q$ = 1.7 \AA$^{-1}$ while the HDL curve (gray) has a peak at $q$ = 2.15 \AA$^{-1}$, which are close to the main peaks in our simulated $I(q)$. (c) The extrapolated structure factor of HDL and LDL in ref~\cite{soper2000structures}. The HDL curve (e) exhibits a peak and a shoulder while the LDL curve (a) shows two well separated peaks between 1$\sim$4 \AA$^{-1}$, which are similar to our simulated $S(q)$.}
\end{center}
\end{figure}

\begin{figure}[htpb]
\begin{center}
\centerline{\includegraphics[width=8.6cm]{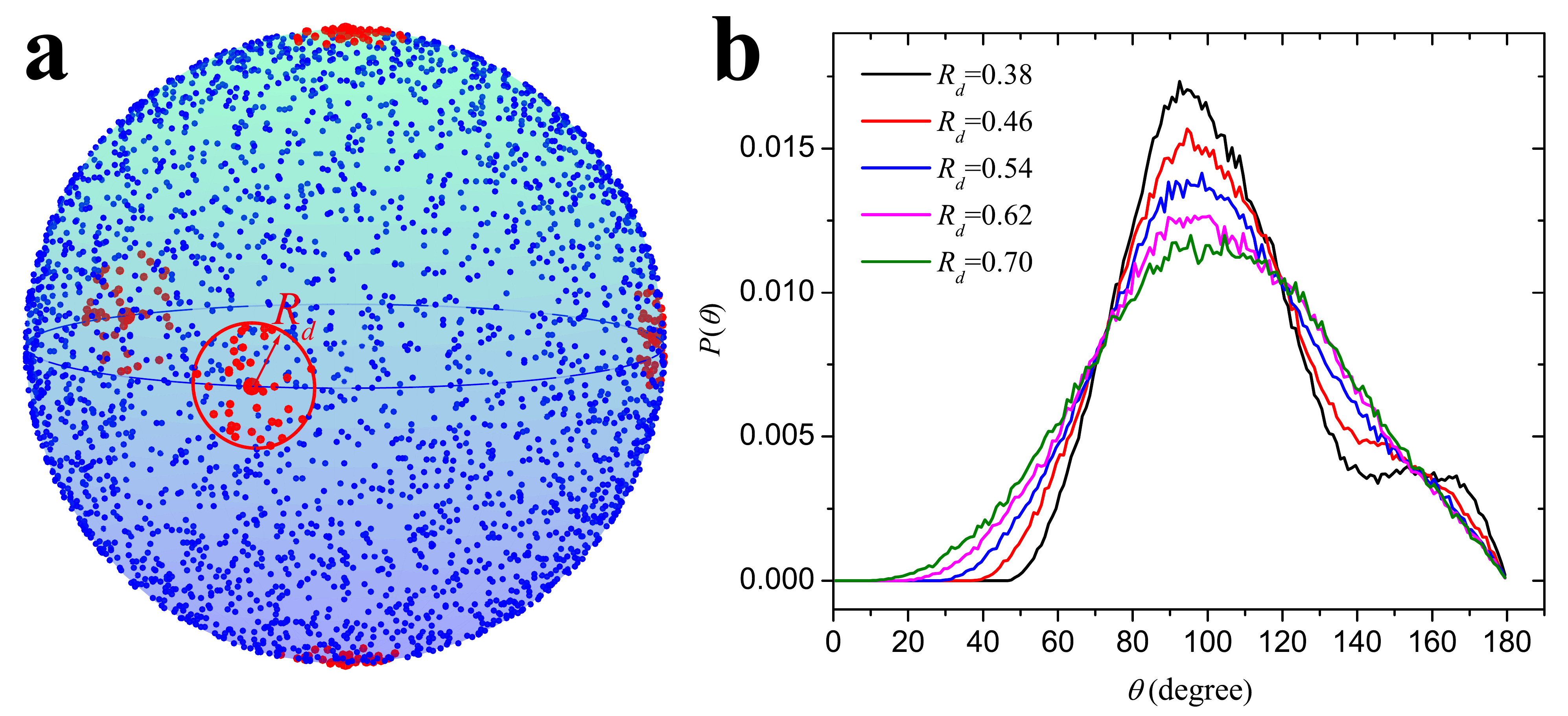}}
\caption{\label{fig:s8}Random perturbations to the uniform distribution of five points on a spherical surface. (a) First we put many ($10^5$) random points uniformly distributed on the spherical surface whose radius is 1. All the points are generated using the method in the ref~\cite{sphereuniform}. The red points represent the points within the distance $R_d$ from the original ideal positions. Five such red points plus the sphere center form one specific configuration mimicing the 5-hydrogen-bond structure. Numerous such configurations ($3*10^4$) are then obtained for good statistics. (b) The distribution P($\theta$) under different $R_d$ perturbations. The distribution agrees well with the simulated 5-hydrogen-bond structure when $R_d$ equals to 0.46 (shown in the main text Fig.~4(d)).}
\end{center}
\end{figure}

\section*{APPENDIX I: Structures of LDL and HDL}
According to the fitting parameters in Eq.~(\ref{S13}), we know that LDL has lower energy and entropy compared with HDL. Therefore, LDL tends to dominate at low temperatures. Following previous research~\cite{russo2014understanding,shi2020direct}, we construct LDL dominant system at $P$ = -2000 bar, $T$ = 200.0 K and HDL dominant system at $P$ = 10000 bar, $T$ = 250.0 K in simulation. Using Eq.~(\ref{S10}), their scattering intensity $I(q)$ can be calculated from simulation, and the corresponding molecular structure factor $S(q)$ is obtained, both are shown in Fig.~\ref{fig:s7}(a). Obviously, the two systems exhibit different features in their $I(q)$ and $S(q)$ curves. In their $I(q)$ curves, the HDL dominant system has one main peak at $q$ = 2.28 \AA$^{-1}$ while the main peak for the LDL dominant system is at 1.68 \AA$^{-1}$. In a recent experiment~\cite{kim2020experimental}, similar peaks at $q$ = 2.15 \AA$^{-1}$ for HDL and $q$ = 1.7 \AA$^{-1}$ for LDL were observed during the liquid-liquid transition process (shown in Fig.~\ref{fig:s7}(b)), which agrees well with our simulation results. In addition, our HDL dominant system's $S(q)$ exhibits a main peak at $q$ = 2.32 \AA$^{-1}$ and a shoulder at $q$ = 2.97 \AA$^{-1}$ while the LDL dominant system exhibits two peaks at $q$ = 1.71 \AA$^{-1}$ and 3.03 \AA$^{-1}$. These properties also agree well with the results in a previous neutron diffraction experiment~\cite{soper2000structures} (shown in Fig.~\ref{fig:s7}(c)). The two well-separated peaks in the $S(q)$ of LDL dominant system indicate the separation of its first and second shells at molecular level. The separation between the two shells results in a low-density and ordered structure with low energy and entropy. By contrast, for the $S(q)$ of HDL dominant system, the main peak and the shoulder-like peak on its right are not well separated, due to the collapse of the first and second shells in the molecular structure. This results in a high-density and disordered structure with high energy and entropy. The agreement between experiment and simulation provides a strong support for our simulation.

\section*{APPENDIX J: Structure of the 5-hydrogen-bond component}
To explore the structure of the 5-hydrogen-bond component in real space, we calculate its angle distribution P($\theta$) as the following: we first connect the central oxygen atom with the five surrounding oxygen atoms to form five straight lines, the angle between any two lines is then defined as $\theta$ and the distribution of these angles gives P($\theta$). P($\theta$) at different temperatures are shown in the main text Fig.~4(d). Obviously, P($\theta$) is not a random distribution and some specific structure should exist. To figure out this structure, we note that all five hydrogen bonds have similar lengths, and approximately assume that the five surrounding oxygen atoms are uniformly distributed on a spherical surface. How to distribute $N$ points uniformly on a spherical surface? This is the so called $Spherical\ Code$ or $Thomson\ problem$. For our $N = 5$ situation, putting two points at north and south pole respectively, and the rest three on the equator with the angle $120^\circ$ is the most uniform configuration~\cite{whyte1952unique,edmundson1992distribution}. To mimic thermal effect, we add random perturbations to the five points' positions as the following. We first generate 10$^5$ points randomly and uniformly distributed on a spherical surface~\cite{sphereuniform}, as shown in Fig.~\ref{fig:s8}(a). Then we find all the points within the distance of $R_d$ from the five ideal points. We randomly choose one point for each center and form one specific configuration and calculate the corresponding angles $\theta$. For each $R_d$, we calculate 3*10$^4$ random configurations to get the angle distribution as shown in Fig.~\ref{fig:s8}(b). We find that the angle distribution agrees the best with the actual simulation results when $R_d$ equals to 0.46 of sphere radius (shown in the main text Fig.~4(d)), which indicates that 0.46 best matches the thermal fluctuations around our simulation temperature range. Therefore, the 5-hydrogen-bond structure can be considered as five water molecules uniformly distributed around the central molecule on a spherical surface. Due to the high symmetry~\cite{edmundson1992distribution} and low potential energy of this configuration, the 5-hydrogen-bond component exhibits a very stable fraction with temperature change, which is different from LDL and HDL.

\section*{APPENDIX K: Differentiating three components and calculating their local densities}
\begin{figure}[htpb]
\begin{center}
\centerline{\includegraphics[width=8.6cm]{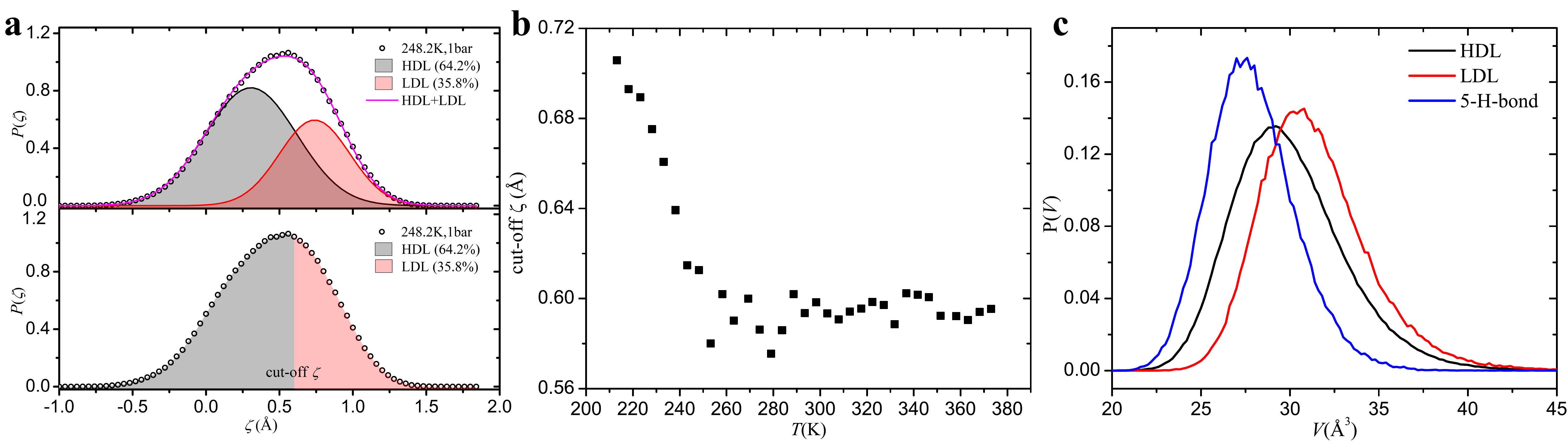}}
\caption{\label{fig:s9}(a) Our approach to differentiate the LDL and HDL at the molecular level. Upper panel: the two-Gaussian decomposing method to calculate the fractions of HDL and LDL. Lower panel: a cut-off $\zeta$ is chosen to separate HDL from LDL. The cut-off $\zeta$ is determined such that the fractions of HDL and LDL matches the two-Gaussian decomposing method. (b) The cut-off $\zeta$ is quite stable above 260K (-13 $^\circ$C) while it changes significantly below this temperatures. (c) The Voronoi cell volume distributions of the three components in one typical sample at 25 $^\circ$C.}
\end{center}
\end{figure}

The fractions of the three components can be calculated by the order parameter distribution (LDL and HDL) and the unique 5-H-bond structure. To calculate their local densities, however, a well-defined criterion must be defined to differentiate the three components microscopically. The 5-H-bond component can be easily picked out because of its unique structure, and a convenient method to differentiate the LDL and HDL in the rest molecules is to set a cut-off order parameter $\zeta$ to satisfy the fractions calculated by the previous two-Gaussian decomposing method (shown in Fig.~\ref{fig:s9}(a)). Water molecules with the order parameter less than the cut-off $\zeta$ belong to the HDL component and the rest belong to the LDL component. This method requires the cut-off $\zeta$ to change with the temperature as shown in Fig.~\ref{fig:s9}(b).

\begin{figure}[htpb]
\begin{center}
\centerline{\includegraphics[width=8.6cm]{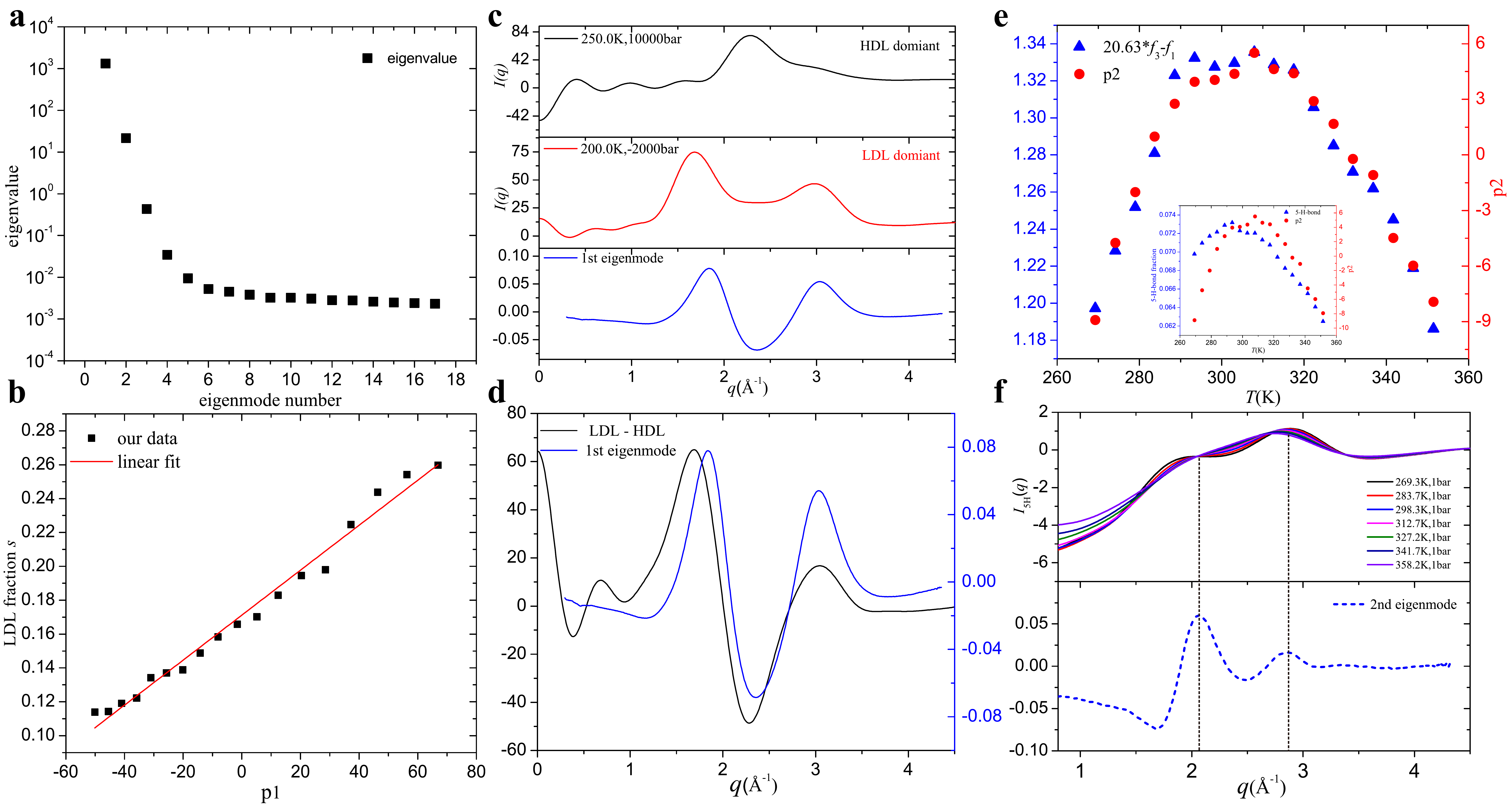}}
\caption{\label{fig:s10}(a) The eigenvalues from large to small. The first two are significantly larger than the rest. (b) LDL fraction has a linear relation with the projection pre-factor $p1$. (c) Top to bottom panels show HDL system's $I(q)$, LDL system's $I(q)$ and the first eigenmode. (d) The first eigenmode agrees well with the curve obtained by the LDL curve minus the HDL curve. (e) Inset: the 5-H-bond fraction $f_3$ from simulation and the projection pre-factor p2 from experiment
are both non-monotonic and having similar proile. Main panel: the non-monotonic $p2$ matches $20.63 \times f_3 - f_1$ nicely. (f) The
main peaks in the second eigenmode correspond nicely to the isosbestic point and the main peak of $I_{5H}(q)$.}
\end{center}
\end{figure}

\begin{figure*}[htpb]
\begin{center}
\centerline{\includegraphics[width=17.2cm]{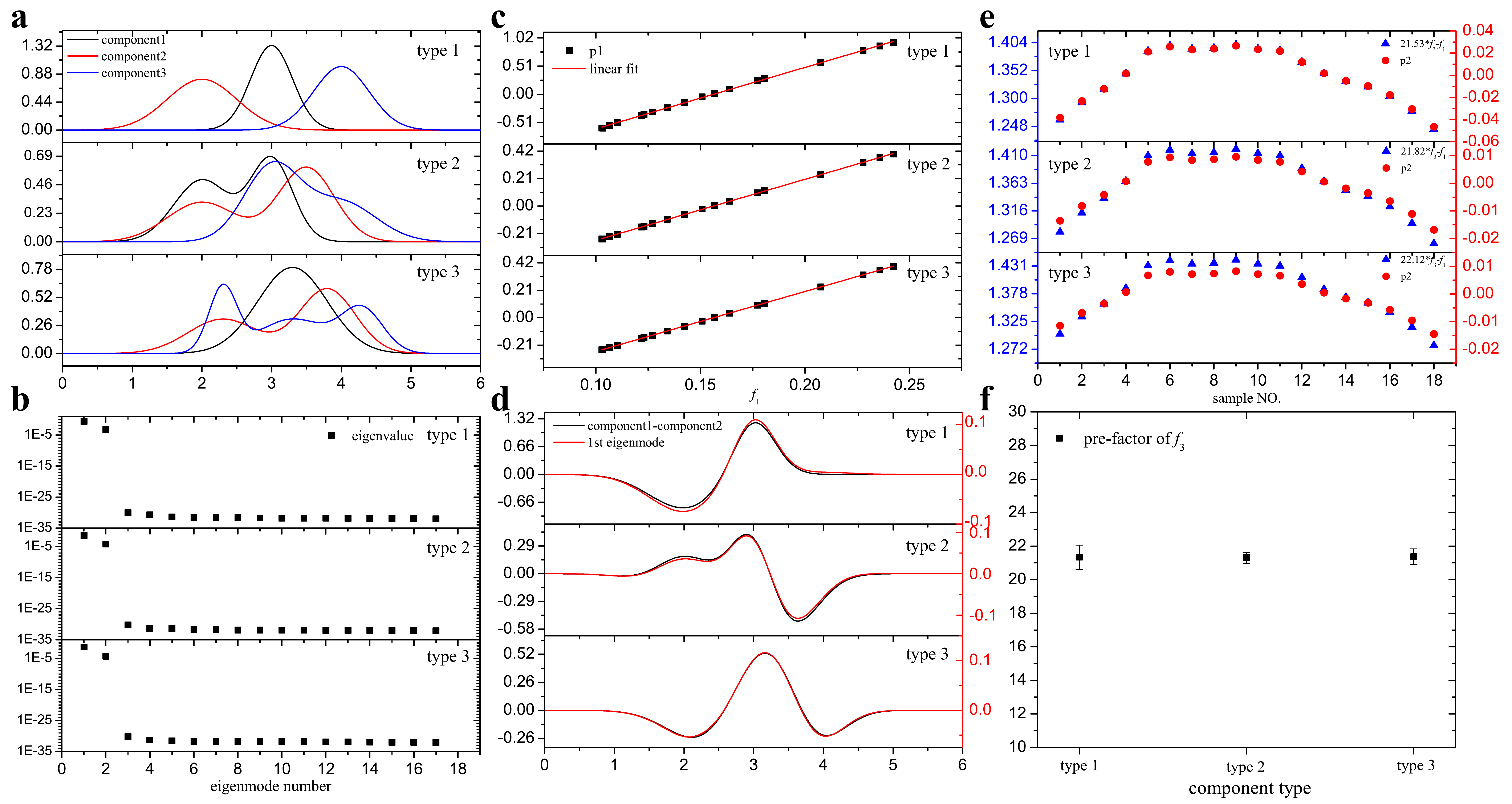}}
\caption{\label{fig:s11}(a) Three different types of the basic components used to generate the data set. Top panel: all the three components are single gaussian functions. Middle panel: all the three components are the sum of two gaussian functions. Bottom panel: the first component is one guassian function, the second component is the sum of two guassian functions, and the third component is the sum of three guassian functions. (b) The eigenvalues of the three different data sets. Obviously, there are only two large eigenvalues. (c) All the three different data sets have a nice linear relationship between $p1$ and $f_1$. (d) The difference between two dominant components overlap well with the 1st eigenmode for all the three types of systems. (e) The linear combinations of $f_1$ and $f_3$ reproduce the same profile as $p2$, as predicted by Eq.~(\ref{eq:3}) in Appendix A, and the pre-factors of $f_3$ are around 21. (f) For more randomly generated data sets, the pre-factor of $f_3$ are also around 21.}
\end{center}
\end{figure*}

Then we calculated the Voronoi cell volume of each water molecule based on the positions of the oxygen atoms. For the oxygen atoms at the boundary of simulation box, their neighboring oxygen atoms should be counted based on the periodic boundary condition, and then their Voronoi cells are constructed. The Voronoi cell calculation is performed with the software Matlab~\cite{matlabVcell}. The distribution of the Voronoi cell volume is shown in Fig.~\ref{fig:s9}(c), and apparently the third component has the smallest volume while the LDL component has the largest volume. Their densities are calculated by their mean Voronoi cell volumes, as shown in the main text Fig. 4(f): The 5-H-bond or 3rd component exhibits an ultra-high density significantly larger than both HDL and LDL.

\section*{APPENDIX L: PCA on $I(q)$ curves for water system}
In general the XRD scattering intensity $I(q)$ is the quantity obtained in actual scattering experiments, which has contributions from both intramolecular and intermolecular correlations, as Eq.~(\ref{S10}) shows for the water system. The relationship between the normalized intensity $I^{nor}(q)$ and $S(q)$ is described by the formula $S(q) - 1 = [I^{nor}(q) - {F^2}(q)]/[2{f_H}(q) + {f_O}(q)]^2$. We note that in our water samples the normalized intensity $I^{nor}(q)$ and $S(q)$ are linearly correlated because the factors $F(q)$, $f_H(q)$ and $f_O(q)$ in the formula are only related to the structure of one single water molecule and thus they keep unchanged with temperature and component variations. Therefore, in water system all the results of PCA on $I(q)$ curves should be equivalent to $S(q)$ curves. We can directly perform PCA on water's $I(q)$ curves at different temperatures and compare the results with $S(q)$ counterparts. First, there are two large eigenvalues meaning that there are two main reasons for water structure's evolution at different temperatures, as show in Fig.~\ref{fig:s10}(a). The dominant reason is mutual conversion between HDL and LDL, which results in the linear relationship between the LDL fraction $s$ and p1 (the projection pre-factor to the 1st eigenmode), as show in Fig.~\ref{fig:s10}(b). Second, according to the analysis in the main text, the 1st eigenmode should correspond to the difference of HDL and LDL, and the results of PCA on $I(q)$ curves also support this interpretation, as show in Fig.~\ref{fig:s10}(c) and (d). Third, the 2nd eigenvalue is related to the fraction change of the 3rd component, 5-H-bond structure. Therefore, $p2$ directly correlates to the linear combination of the fractions, $f_3$ and $f_1$: $p2 \propto 20.63 \times f_3 - f_1$, as show in Fig.~\ref{fig:s10}(e). This formula is identical to the one for $S(q)$'s PCA analysis in the main text and quantitatively proves the equivalence of $I(q)$ and $S(q)$ results. At last, the intensity curves calculated from the 5-H-bond molecules as origins also exhibit main features that match the 2nd eigenmode, as shown in Fig.~\ref{fig:s10}(f). This agrees with the $S(q)$ result in Fig.~4(b) in the main text. To conclude, all the reproducible results unambiguously show that the PCA analysis of $I(q)$ and $S(q)$ curves in our water samples are equivalent.

\section*{APPENDIX M: A general model}
As our PCA analysis on water experiment and three-alcohol simulation reveals reproducible results for a mixture of three components, we would like to test these results in a more general system. Because in general many features are peaks or valleys which can be approximated by Gaussian peaks, without loss of generality, we use Gaussian peaks or their superpositions as basic components, and then add three such components together to form a three-component mixture. A few types of basic components are shown in Fig.~\ref{fig:s11}(a), which are either Gaussian peaks (type 1) or their superpositions (type 2 and 3). To compare with water, we use the identical fractions as water's three components shown in the main text Fig.~4(e), and construct general three-component mixtures with these Gaussian peaks. Applying PCA method to these mixture systems, we can reproduce all results from the water system. First, there are only two large eigenvalues as shown in Fig.~\ref{fig:s11}(b). This indicates that there are two main reasons that cause the system evolution. The first reason is due to the mutual conversion between the two dominant components, as proved by the nice linear relationship between $p1$ and $f_1$ in Fig.~\ref{fig:s11}(c), and the overlap of the first eigenmode and the difference between the two dominant components in Fig.~\ref{fig:s11}(d). The second reason is related to the variation in the third component revealed by the second eigenmode. Once again, we observe a nice match between $p2$ and the linear combination of $f_3$ and $f_1$ in Fig.~\ref{fig:s11}(e), which is similar to water and three-alcohol systems. Moreover, the pre-factor of $f_3$ is also very close to water and three-alcohol systems, regardless of the very different basic components' shapes. To verify this further, we change the Gaussian peaks' height and width randomly for each component type in Fig.~\ref{fig:s11}(a) and calculate the corresponding pre-factors of $f_3$. We find that all the pre-factors are around 21, as shown in Fig.~\ref{fig:s11}(f), which agrees excellently with water's value of 20.63. Therefore, the quantitative agreement of this pre-factor in water, three-alcohol system, and the Gaussian peak system provides unambiguous evidence that our PCA analysis is robust and universal.

\end{document}